\documentclass[10pt,a4paper]{scrartcl}

\usepackage{CLICdp}
\usepackage{amssymb}
\usepackage{CLICdp_definitions}

\usepackage{float}
\usepackage[textwidth=1.5cm]{todonotes}
\usepackage{adjustbox}
\usepackage{sansmath}
\usetikzlibrary{positioning}
\usepackage{threeparttable}
\usepackage{multirow}
\usepackage{makecell}
\usepackage{etoolbox}
\usepackage{afterpage}
\usepackage{placeins}
\usepackage{multicol}
\robustify\bfseries
\usepackage{wrapfig}

\DeclareSIUnit\years{years}
\DeclareSIUnit\days{days}

\title{\hspace*{1cm} The Compact Linear e$^+$e$^-$ Collider (CLIC):\newline Physics Potential}

\notitlestamp 
\date{\formatdate{19}{12}{2018}}

\addauthor{Contact person: P.~Roloff\thanks{philipp.roloff@cern.ch}}{\institute{1}\\}
\addauthor{Editors: R.~Franceschini}{\institute{2}\institute{3},}
\addauthor{P.~Roloff}{\institute{1},}
\addauthor{U.~Schnoor}{\institute{1},}
\addauthor{A.~Wulzer}{\institute{1}\institute{4}\institute{5}}

\addinstitute{1}{CERN, Geneva, Switzerland}
\addinstitute{2}{Universit\`{a} degli Studi Roma Tre, Rome, Italy}
\addinstitute{3}{INFN, Sezione di Roma Tre, Rome, Italy}
\addinstitute{4}{Universit\`{a} di Padova, Padova, Italy}
\addinstitute{5}{LPTP, EPFL, Lausanne, Switzerland}

\onbehalfof{ {\em Input to the European Particle Physics Strategy Update} \\ \vspace*{0cm} {\em on behalf of the CLIC and CLICdp Collaborations} \\ \vspace*{.4cm} {\large 18 December 2018} }

\abstract{The Compact Linear Collider, CLIC, is a proposed $\epem$ collider at the \TeV scale whose
physics potential ranges from high-precision measurements to extensive direct sensitivity to physics beyond the Standard Model.
This document summarises the physics potential of CLIC, obtained in detailed studies, many based on full simulation of the CLIC detector.
CLIC covers one order of magnitude of centre-of-mass energies from 350\,GeV to 3\,TeV, giving access to large event samples for a variety of SM processes, many of them for the first time in $\epem$ collisions or for the first time at all.
The high collision energy combined with the large luminosity and clean environment of the $\epem$ collisions enables the measurement of the properties of Standard Model particles, such as the Higgs boson and the top quark, with unparalleled precision.
CLIC might also discover indirect effects of very heavy new physics by probing the parameters of the Standard Model Effective Field Theory with an unprecedented level of precision.
The direct and indirect reach of CLIC to physics beyond the Standard Model significantly exceeds that of the HL-LHC. This includes new particles detected in challenging non-standard signatures.
With this physics programme, CLIC will decisively advance our knowledge relating to the open questions of particle physics. 
}

\graphicspath{ {./logos/}{./figures/} }

\usepackage{cleveref}
\newlength{\abc}
\AtBeginDocument{%
\renewcommand{\ref}[1]{\mbox{\Cref{#1}}}

}

\begin{document}

\titlepage

\section{Introduction}\label{sec:Intro}
The Compact Linear Collider (CLIC) is a multi-TeV high-luminosity linear $\epem$ collider under development~\cite{clic-study,ESU18Summary}. CLIC uses a novel two-beam acceleration method, with normal-conducting structures operating with gradients in the range of 70--100\,MV/m. This approach provides the only mature option for a multi-TeV lepton collider. In parallel, a detector to study the $\epem$ collisions at CLIC is being designed based on a broad range of simulation studies and R\&D on crucial technological aspects~\cite{CLICdet_note_2017, CLICdet_performance}.

To cover the widest possible range of physics opportunities, CLIC will be constructed in several centre-of-mass energy stages. A first stage at 380\,GeV gives access to the Higgsstrahlung process, which in $\epem$ collisions enables a unique Higgs physics programme, and to the top quark, which so far has only been produced in hadron collisions. The higher-energy stages, currently assumed to be at 1.5\,\TeV and 3\,\TeV, provide unique sensitivity for a large number of new physics senarios and give access to the Higgs self-coupling.

The current CLIC baseline staging scenario~\cite{Roloff:2645352}, based on accelerator ramp-up and up-time scenarios harmonised with those of other potential future colliders, assumes 1.0\,ab$^{-1}$, 2.5\,ab$^{-1}$ and 5.0\,ab$^{-1}$ of luminosity collected at the three energy stages of $\sqrt{s}$ = 380\,GeV, 1.5\,TeV and 3\,TeV, respectively. The first stage includes 100\,fb$^{-1}$ obtained in an energy scan around the $\PQt\PAQt$ production threshold. The complete physics programme will span 25--30 years. The baseline CLIC accelerator provides $\pm80\%$ longitudinal polarisation for the electron beam, and no positron polarisation. Equal amounts of $-80\%$ and $+80\%$ polarisation running are foreseen at the initial energy stage. At the two higher-energy stages, a sharing of the running time for $-80\%$ and $+80\%$ electron polarisation in the ratio $80:20$ is assumed. The baseline scenario as described above is summarised in \ref{tab:clic_stages}.

\begin{table*}[htp]\centering
\caption{Baseline CLIC energy stages and integrated luminosities for each stage in the updated scenario. \label{tab:clic_stages}}
\begin{tabular}{ccc|cc}\toprule
 & & & $P(\Pem)=-80\%$ & $P(\Pem)=+80\%$ \\
Stage & $\sqrt{s}$ [TeV] & $\mathcal{L}_\mathrm{int}$ [ab$^{-1}$] & $\mathcal{L}_\mathrm{int}$ [ab$^{-1}$] & $\mathcal{L}_\mathrm{int}$ [ab$^{-1}$] \\
\hline
1 &  0.38 (and 0.35) &  1.0 & 0.5 & 0.5 \\
2 &  1.5             &  2.5 & 2.0 & 0.5 \\
3 &  3.0             &  5.0 & 4.0 & 1.0 \\
\bottomrule
\end{tabular}
\end{table*}

The CLIC physics programme will enable fundamentally new insights beyond the capabilities of the HL-LHC. The flexibility and large accessible energy range, almost one order of magnitude, provides a wide range of possibilities to discover new physics using very different approaches. The high centre-of-mass energy of CLIC extends the direct mass reach to scales far greater than that available at previous lepton colliders, surpassing even the HL-LHC for many signatures. The high luminosity and absence of QCD backgrounds give access to very rare processes at all energies. The clean experimental environment and absence of triggers in high-energy $\epem$ collisions and the good knowledge of the initial state allow precise measurements of many reactions to be performed, which probe the effects of new physics at mass scales far beyond the kinematic reach for direct production of new particles. The use of electron beam polarisation enhances this reach further and may help to characterise newly discovered phenomena. Threshold scans provide very precise measurements of known particle masses. The CLIC experimental environment is also well-suited for looking for non-standard signatures such as anomalous tracks, peculiar secondary vertices, or unexpected energy depositions in the calorimeters.

These capabilities make CLIC an ideal option for the next large facility in high energy physics. The first energy stage at 380\,GeV provides an exciting programme of precision Higgs and top-quark physics while the energies of the succeeding stages can be adapted to possible input from the HL-LHC or earlier CLIC running. In the following, we will discuss the physics potential of CLIC from two different perspectives. \ref{sec:PrecisionSM} explores important Standard Model (SM) processes with an emphasis on Higgs boson and top-quark physics. First conclusions on the indirect sensitivity to new physics scales are obtained from Effective Field Theory. \ref{sec:DirectBSM} shows how this indirect reach, combined with the great direct exploration potential, allows CLIC to make decisive progress on a number of concrete Beyond the Standard Model (BSM) questions and scenarios.

The CLIC accelerator and detector are described in a separate submission to the European Strategy Update process, `The Compact Linear \epem Collider (CLIC): Accelerator and Detector'~\cite{ESU18project}; this and supporting documents can be found at the following location:

\begin{center}
\textbf{Supporting documents: \url{http://clic.cern/european-strategy}}
\end{center} 

\section{Learning from Standard Model processes}\label{sec:PrecisionSM}
A key component of the CLIC physics programme studies the production and decay properties of the known SM particles. The experimental conditions at CLIC allow the selection of many relevant final states with high efficiency while keeping the background contributions at a low level. The known centre-of-mass energy provides an additional kinematic constraint for event reconstruction that is not available in hadron collisions. In this section, the CLIC potential for Higgs boson and top-quark physics, as well as two-fermion and multi-boson production processes, is discussed.

\subsection{Higgs boson}
After the discovery of a Higgs boson, the investigation of its properties and with it the nature of the mechanism of electroweak symmetry breaking has top priority in particle physics.
New physics scenarios in the Higgs sector include extended Higgs sectors, composite Higgs, and models with the Higgs boson as a portal to a BSM sector.

\begin{wrapfigure}{r}{0.4\textwidth}
\begin{center}
\vspace{-20pt} 
\includegraphics[width=\linewidth]{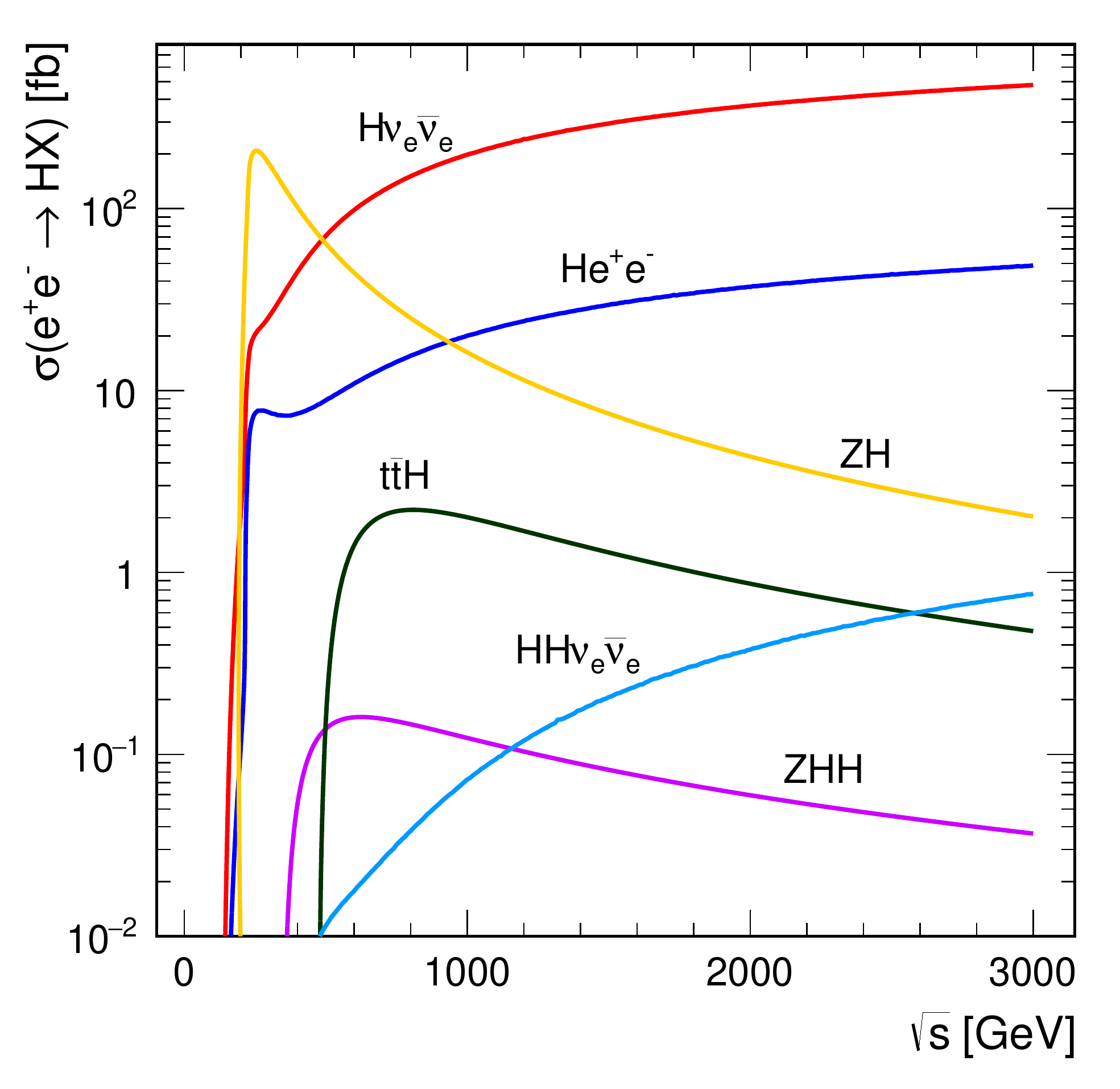}
\caption{Production cross section as a function of centre-of-mass energy for the main Higgs boson production processes
at an \epem collider~\cite{ClicHiggsPaper}.}
\label{fig:higgs_xs_energy}
\end{center}
\vspace{-20pt}
\end{wrapfigure}

\ref{fig:higgs_xs_energy} shows the centre-of-mass energy dependence of the relevant Higgs boson production processes in
$\epem$ interactions. The Higgsstrahlung mechanism, $\epem\to\PZ\PH$, is dominant at 380\,GeV.
This process can be identified using the mass recoiling against the $\PZ$ boson, which provides the Higgs branching ratios and
width in a model-independent manner, without any assumptions about BSM invisible decays; a feature that is unique to lepton colliders.
In total, 160000 Higgs bosons are produced at the first stage of CLIC operation. At 1.5 and 3 TeV, large Higgs boson samples are produced
in $\PW\PW$ fusion, $\epem\to\PH\PGne\PAGne$. The full CLIC baseline scenario provides about 4.5 million Higgs boson decays.
The Higgs self-coupling can be probed in  $\epem\to\PZ\PH\PH$ and $\epem\to\PH\PH\PGne\PAGne$ events
at high energy. The CLIC Higgs reach has been comprehensively investigated using full simulation studies~\cite{Roloff:2645352, ClicHiggsPaper}.
The CLIC projections for the extraction of the Higgs couplings and the Higgs self-coupling are summarised in the following\footnote{The Higgs physics projections
were obtained assuming slightly different energies for the first two stages: 350\,GeV instead of 380\,GeV and 1.4\,TeV instead of 1.5\,TeV.}.

\begin{figure}[bpht]
\centering
\begin{subfigure}{.4\textwidth}
\includegraphics[width=\linewidth]{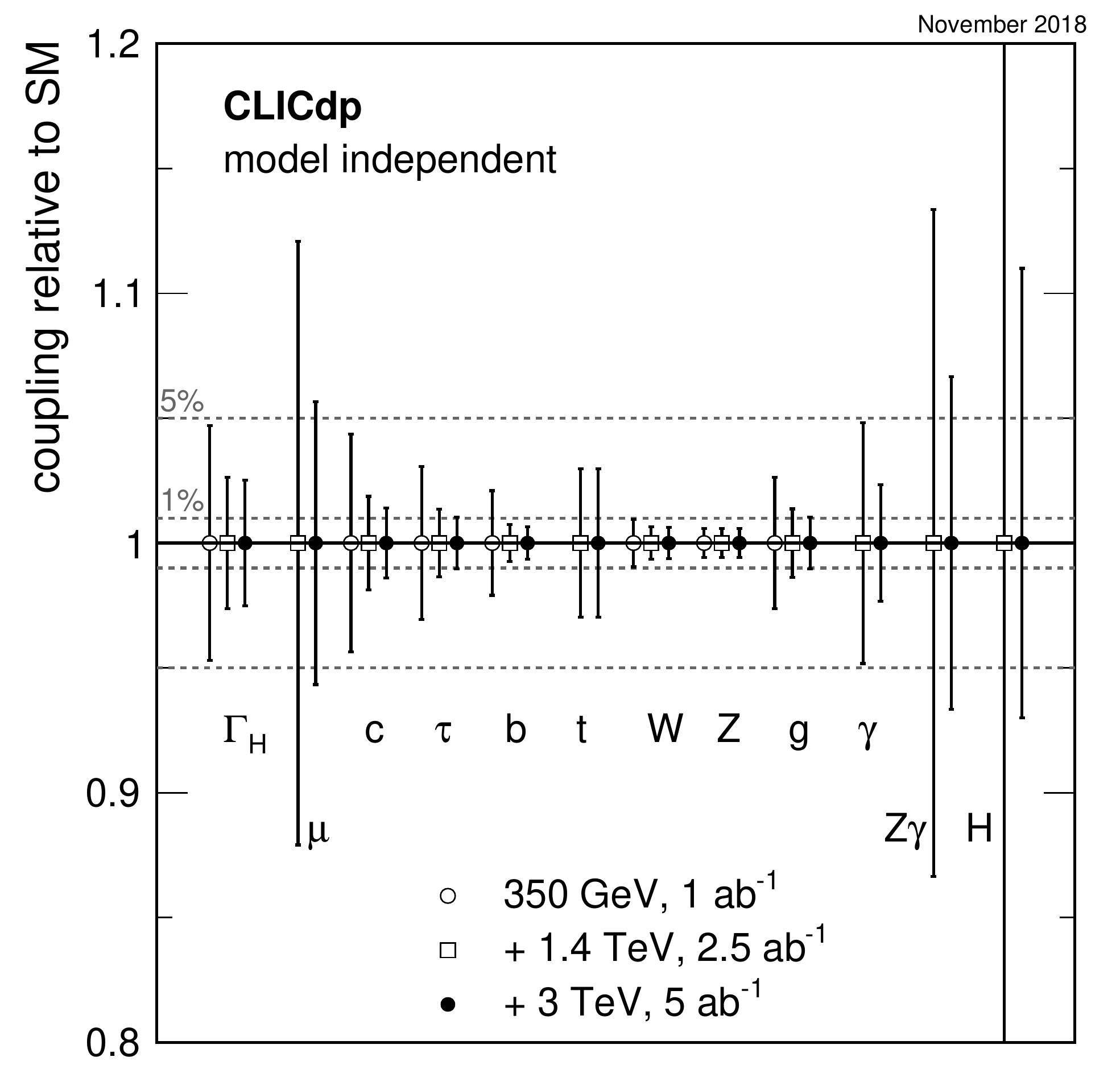}\label{fig:MIResultsPolarised8020_esu}
\end{subfigure}
\begin{subfigure}{.4\textwidth}
\includegraphics[width=\linewidth]{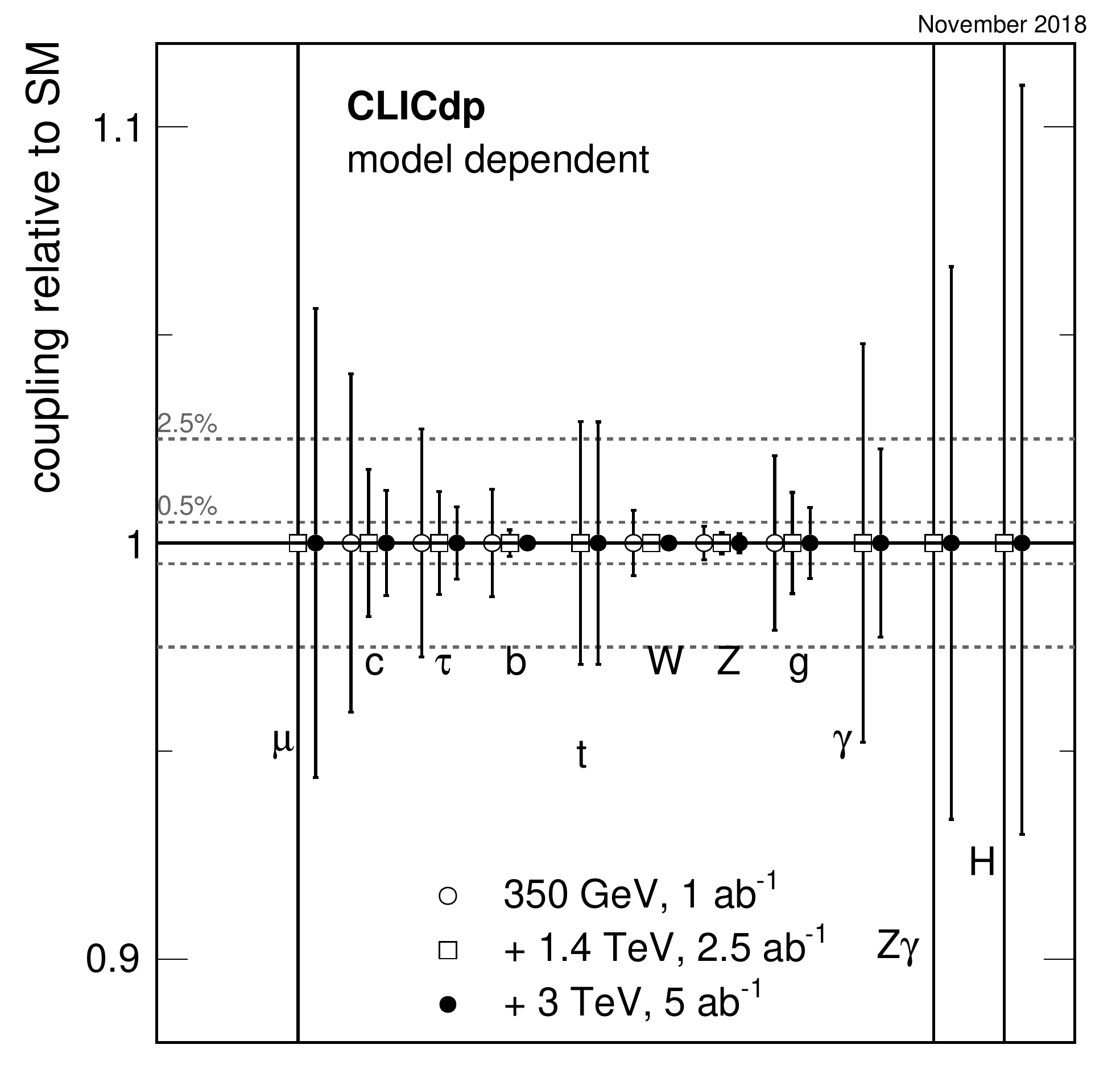}\label{fig:MDResultsPolarised8020_esu}
\end{subfigure}\vspace{-5pt}
\caption{CLIC results of (left) the model-independent fit and (right) the model-dependent fit to the Higgs couplings to SM particles~\cite{Roloff:2645352,ClicHiggsPaper}. For the top-Higgs coupling, the $\SI{3}{\TeV}$ case has not yet been studied.}
\label{fig:HiggsResultsPolarised8020_esu}
\vspace{-10pt}\end{figure}

\paragraph{Higgs couplings}
The measurements of Higgs boson production cross sections times branching fractions in all accessible channels
can be combined to extract the Higgs couplings and width. A model-independent global fit, described
in \cite{ClicHiggsPaper}, makes use of the total cross section for the Higgsstrahlung process measured using the
recoil mass method at the first energy stage to avoid any assumptions about additional BSM decays. For this reason, the initial CLIC
stage is crucial for the Higgs physics programme. The results of the fit for the three CLIC energy stages are shown
in \ref{fig:HiggsResultsPolarised8020_esu}(left). The expected precision on \gHZZ is 0.6\,\% from the total $\PZ\PH$ cross section.
Other couplings such as \gHWW and \gHbb reach similar precision in the model-independent approach. The \gHcc coupling, which
is very challenging at hadron colliders, can be probed with percent-level precision. The total Higgs width is extracted with
2.5\% accuracy. The Higgsstrahlung process at the first CLIC stage can also be used to set a model-independent upper limit on invisible 
Higgs boson decays of BR(H\,$\to$\,invis.)$<$ 0.69\% at 90\% C.L. from the recoil mass spectrum.

Results from a global fit under the assumption of no non-SM Higgs boson decays, which is model-dependent and equivalent 
to the approach at hadron colliders, are illustrated in \ref{fig:HiggsResultsPolarised8020_esu}(right). In this case, several
Higgs couplings are constrained to per mille-level precision at the high-energy stages.

\paragraph{Higgs self-coupling}
The Higgs self-coupling deserves special attention as the HL-LHC~\cite{Gori:2650162} is not sensitive to its SM value.
Deviations from the SM expectation can reach tens of percent in various new physics scenarios.
Double-Higgs boson production accessible at high-energy CLIC operation is sensitive to 
the Higgs self-coupling $\lambda$ at tree level. The second energy stage allows a $5\,\upsigma$-observation
of the double Higgsstrahlung process $\epem\to\PZ\PH\PH$ and provides evidence for the  $\PW\PW$ fusion process 
$\epem\to\PH\PH\PGne\PAGne$ with a significance of $3.6\,\upsigma$ assuming the SM value of $\lambda$. Both measurements are complementary as the 
$\PZ\PH\PH$ ($\PH\PH\PGne\PAGne$) cross section increases (decreases) with $\lambda$ in the region around the SM value.

At $\roots=\SI{3}{\TeV}$, $\PW\PW$ fusion is the leading double-Higgs boson production mechanism.
The cross section is large enough to enable the measurement of differential cross sections to improve the knowledge of the
Higgs self-coupling further. Using both processes at the second stage and differential distributions for $\PH\PH\PGne\PAGne$ production at 
3\,TeV leads to an expected precision on the Higgs self-coupling $\lambda$ of  $[-7\%, +11\%]$~\cite{ESU18BSM}.
The inclusion of the $\PZ\PH\PH$ cross section and the use of differential distributions avoids an ambiguity that occurs in the
extraction of $\lambda$ from the $\epem\to\PH\PH\PGne\PAGne$ cross section alone. 
Given the precision that can be achieved and the possible sizes of deviations in relevant extensions of the SM, the Higgs self-coupling measurement is 
a very important motivation for CLIC operation in the multi-TeV region.

\subsection{Top quark}

\begin{wrapfigure}{R}{0.45\textwidth}
\begin{center}
\vspace{-34pt}
\includegraphics[width=\linewidth]{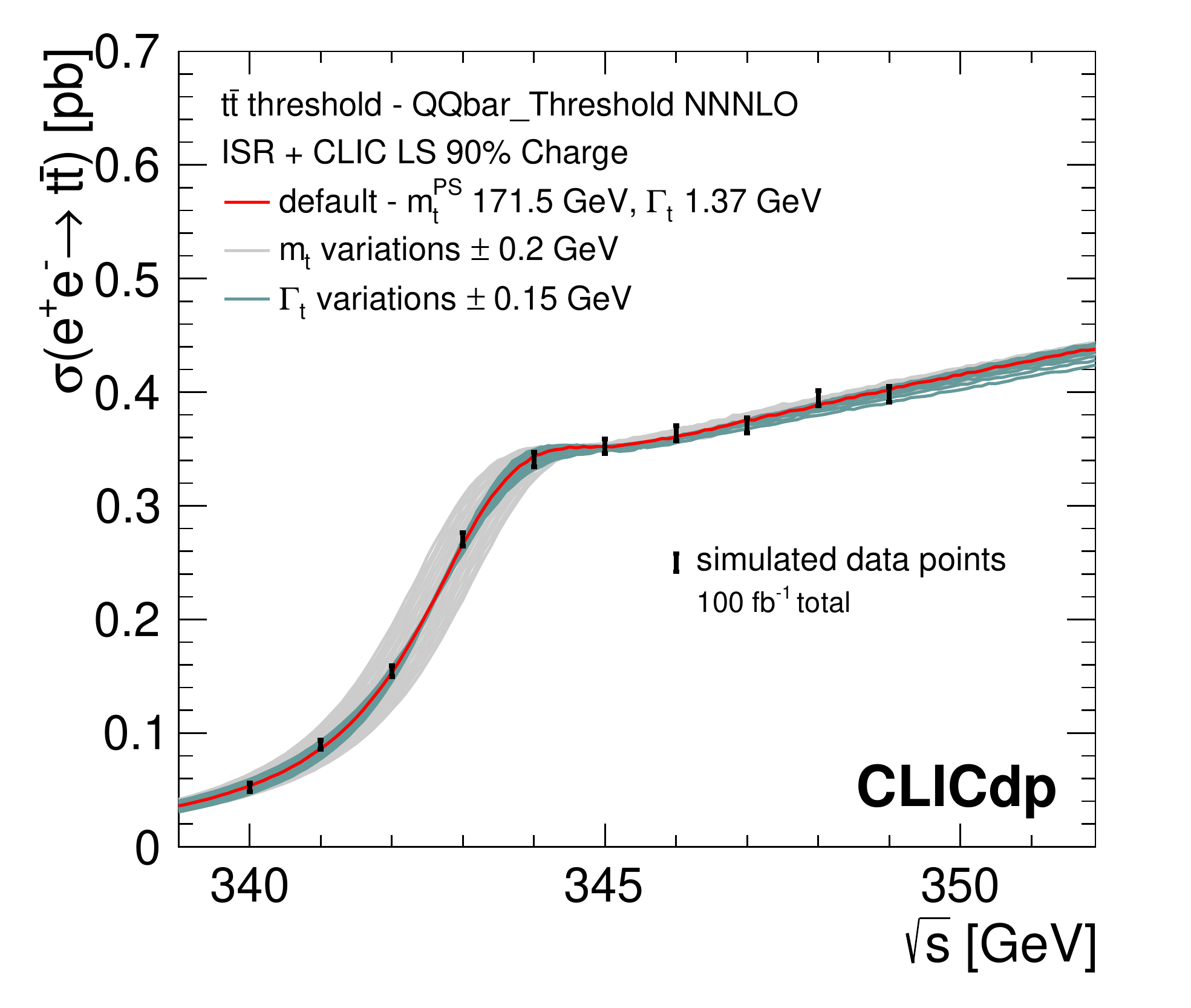}
\caption{Illustration of a top-quark threshold scan at CLIC with a total integrated luminosity
of \SI{100}{\per\fb}, with an optimised reduced bunch charge scenario~\cite{ClicTopPaper}.
The bands around the central cross section curve show the dependence of the cross section on
the top-quark mass and width, illustrating the sensitivity of the threshold scan. The error bars
on the simulated data points show the statistical uncertainties of the cross section measurement.}
\label{fig:TopThresholdScan_esu}
\vspace{-34pt}
\end{center}
\end{wrapfigure}

The top quark is the heaviest known fundamental particle and plays an important role in
many BSM theories; it therefore provides unique opportunities to test the SM and probe signatures of BSM effects.
Already the first CLIC stage provides an important set of measurements using the $\epem\to\PQt\PAQt$ process.
A theoretically well-defined top-quark mass measurement can be performed in a threshold scan. The pair production and decay
of the top quark can be studied at 380\,GeV. The higher-energy stages provide complementary information on
$\PQt\PAQt$ production. Additionally, high-energy CLIC operation gives access to the
$\PQt\PAQt\PH$ final state and to top-quark pair-production in vector boson fusion.
The CLIC prospects for top-quark physics have been studied with full detector simulation~\cite{ClicTopPaper}.
In the following, the highlights of the top-physics programme at CLIC are summarised\footnote{The top-physics projections
were obtained assuming a slightly different energy for the second stage: 1.4\,TeV instead of 1.5\,TeV.}.

\paragraph{Threshold scan}
The top-quark mass can be measured in a dedicated scan collecting \SI{100}{\per\fb} over several centre-of-mass energy values around the top-quark pair-production threshold. This method allows the extraction of a theoretically well-defined mass, which reduces the overall uncertainty significantly.

A highly pure sample of top-quark events can be selected and used to measure the cross section at each energy point. 
The top-quark mass is extracted from the evolution of the cross section as a function of \roots\ as 
shown in \ref{fig:TopThresholdScan_esu}. At CLIC, a statistical uncertainty on the top-quark mass of \SI{20}{\MeV} 
can be achieved. The total uncertainty of around \SI{50}{\MeV} is currently dominated by the scale uncertainties of the 
NNNLO QCD prediction for the top threshold region. The extracted mass in the so-called PS scheme can be translated to 
the $\overline{\rm MS}$ mass scheme with a small additional uncertainty of about \SI{10}{\MeV}.

\paragraph{Top-quark pair production}
The top-quark couplings to the photon and $\PZ$ boson are precisely predicted by the SM
but may receive substantial corrections from BSM physics; for example, theories with extra dimensions
and compositeness can modify the couplings significantly. Here we demonstrate the new physics
potential of $\PQt\PAQt$ pair production using SM Effective Field Theory (SM-EFT).
SM-EFT extends the SM Lagrangian to include interaction operators of higher dimension.
The leading effects are captured by dimension-6 operators weighted by the coefficients $c_{i}/\Lambda^{2}$ for 
dimensionless couplings $c_i$ and a common suppression scale $\Lambda$. Measurements with different beam polarisation
allow the photon and $\PZ$ boson contributions to be disentangled, while data from two or more different
centre-of-mass energies constrain operators whose effects grow with energy. The clean experimental
environment allows differential distributions of the top-quark production
and decay kinematics to be exploited.

At the first stage, an almost background-free sample of semi-leptonic $\PQt\PAQt$ events can
be obtained due to the excellent jet energy resolution and b-tagging performance of the CLIC detector.
Jet substructure techniques are applied at the higher centre-of-mass energies.

\begin{wrapfigure}{R}{0.6\textwidth}
\begin{center}
\vspace{-20pt}
\includegraphics[width=\linewidth]{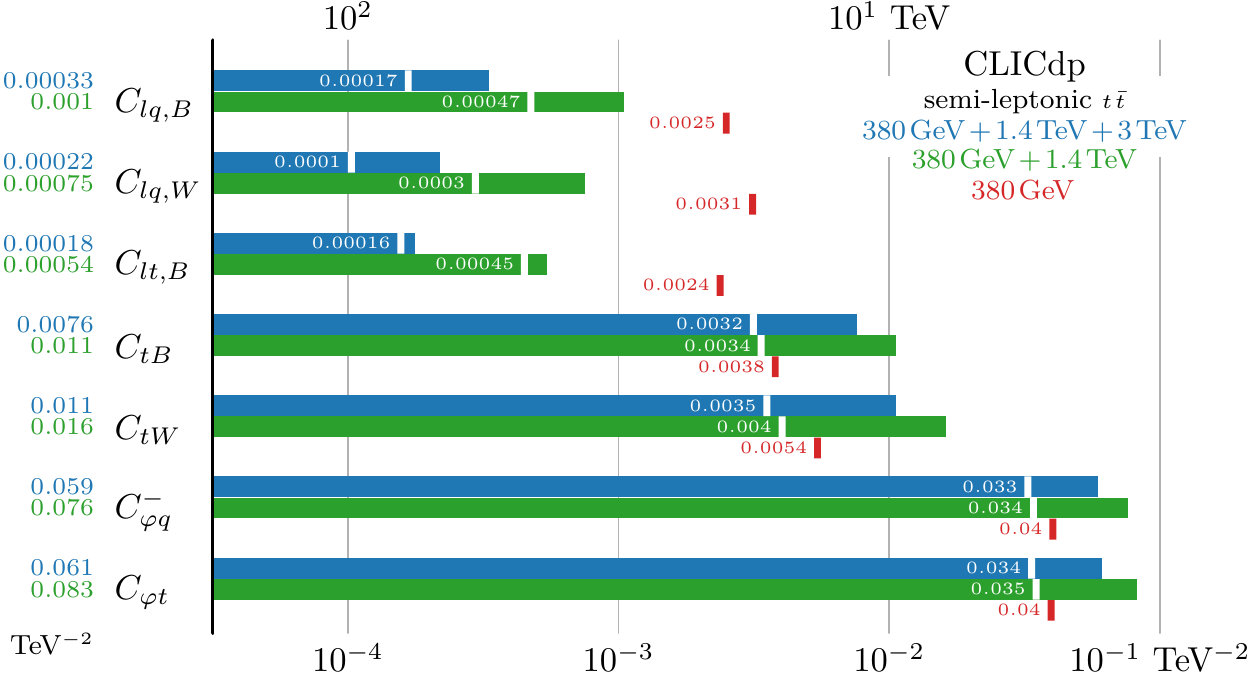}
\caption{EFT analysis of $\PQt\PAQt$ production at CLIC~\cite{ClicTopPaper}. The blue and green bars show
the results of a global fit of seven operator coefficients. The corresponding individual operator
sensitivities are shown as ticks. The sensitivities are measured in TeV$^{-2}$ in the lower horizontal axis,
while the upper horizontal axis gives the sensitivity to the operator scale $\Lambda = 1/\sqrt{C}$.}
\label{fig:TopPhilicEFT_esu}
\vspace{-20pt}
\end{center}
\end{wrapfigure}

The expected global and individual sensitivities for seven SM-EFT operator coefficients (defined in~\cite{ClicTopPaper}) obtained from semi-leptonic
$\PQt\PAQt$ events is shown in \ref{fig:TopPhilicEFT_esu}. The measurements at one single centre-of-mass energy are not sufficient for all operator
coefficients to be extracted simultaneously. At least two stages are needed for a global fit. In general, new physics scales from a few up to tens of 
TeV can be reached. Operation at high energy dramatically improves the sensitivity to the 4-fermion operator 
coefficients $C_{lq,B}$, $C_{lq,W}$ and $C_{lt,B}$, and to a lesser extent, to the dipole operator coefficients
$C_{tB}$ and $C_{tW}$. This result demonstrates the strong benefit of several energy 
stages for the CLIC physics potential.

\paragraph{Associated production with a Higgs boson}
At the second CLIC energy stage, $\PQt\PAQt\PH$ production gives direct access to the top-quark Yukawa coupling.
The top Yukawa coupling can be extracted with a precision of 2.9\,\%~\cite{Roloff:2645352, ClicTopPaper}.
In addition, the $\PQt\PAQt\PH$ process is sensitive to a CP-odd contribution to the $\PQt\PAQt\PH$ coupling.
The mixing of the CP-even and -odd components can be measured to a precision of $\Delta\sin^{2}{\phi} = 0.07$ or better, where $\phi$ is the mixing angle~\cite{ClicTopPaper}.
In the EFT language of \ref{fig:TopPhilicEFT_esu}, the $\PQt\PAQt\PH$ cross section measured at 3 TeV would improve the knowledge
on $C_{tW}$ to the level of 10$^{-3}$\,TeV$^{-2}$~\cite{ESU18BSM}.

\subsection{Other processes: Two-fermion and multi-boson production}
\paragraph{Two-fermion production and electroweak precision tests}
Processes of two-fermion to two-fermion scattering are a sensitive probe to new physics.
In addition to top-quark pair-production described above, CLIC can investigate charged lepton-pair as well as $\cc$ and $ \bb$ production.
As the contribution from four-fermion interactions to the total cross section grows with centre-of-mass energy, 
the combination of high energy and clean environment at CLIC makes it an ideal choice for these measurements.
Deviations from the SM in sensitive observables such as the total cross section and the polar scattering angle in two-fermion production are described within the simplified context of universal new physics by the ``oblique parameters'' \textit{S}, \textit{T}, \textit{W}, and \textit{Y}.
While the effects described by the \textit{S} and \textit{T} parameters are constant with energy, those of the \textit{W} and \textit{Y} parameters grow.
The sensitivities of CLIC to the \textit{S} and \textit{T} parameters have been shown to be similar to current measurement uncertainties from  electroweak precision observables.
The precision of \textit{W} and \textit{Y} measurements, however, will be substantially improved at CLIC by around three orders of magnitude compared with LEP, and one order of magnitude compared with HL-LHC projections~\cite[Sec.~2.6]{ESU18BSM}. 

\paragraph{Multi-boson production and vector boson scattering}
Processes involving two or more electroweak gauge bosons can give hints about new physics in the electroweak sector.
BSM amplitudes are distinguished from the SM amplitudes by their behaviour in sensitive observables such as azimuthal and polar decay angles in di-boson production~\cite[Sec.~2.4]{ESU18BSM}.
In the  $\Pep\Pem \to \PWp\PWm$ process at CLIC, the hadronic decay channel enables the full kinematic reconstruction of the  final state.
Semi-leptonic final states are advantageous for observables that distinguish between the charges of the reconstructed $\PW$ bosons. The momentum of the neutrino can be determined better in $\epem$ collisions than at hadron colliders.
As a result, the limits on anomalous triple gauge couplings from CLIC will significantly improve those expected for the HL-LHC.
CLIC will furthermore enable inclusive measurements of the production of two and more electroweak gauge bosons in annihilation and vector boson scattering (VBS)~\cite[Sec.~2.5]{ESU18BSM}. For well-constrained dimension-6 EFT operators, tri-boson production and VBS give strong constraints on dimension-8 operator coefficients.

\subsection{Effective field theory fit and interpretation}

\begin{figure}[htbp]
\begin{center}
\includegraphics[width=\linewidth]{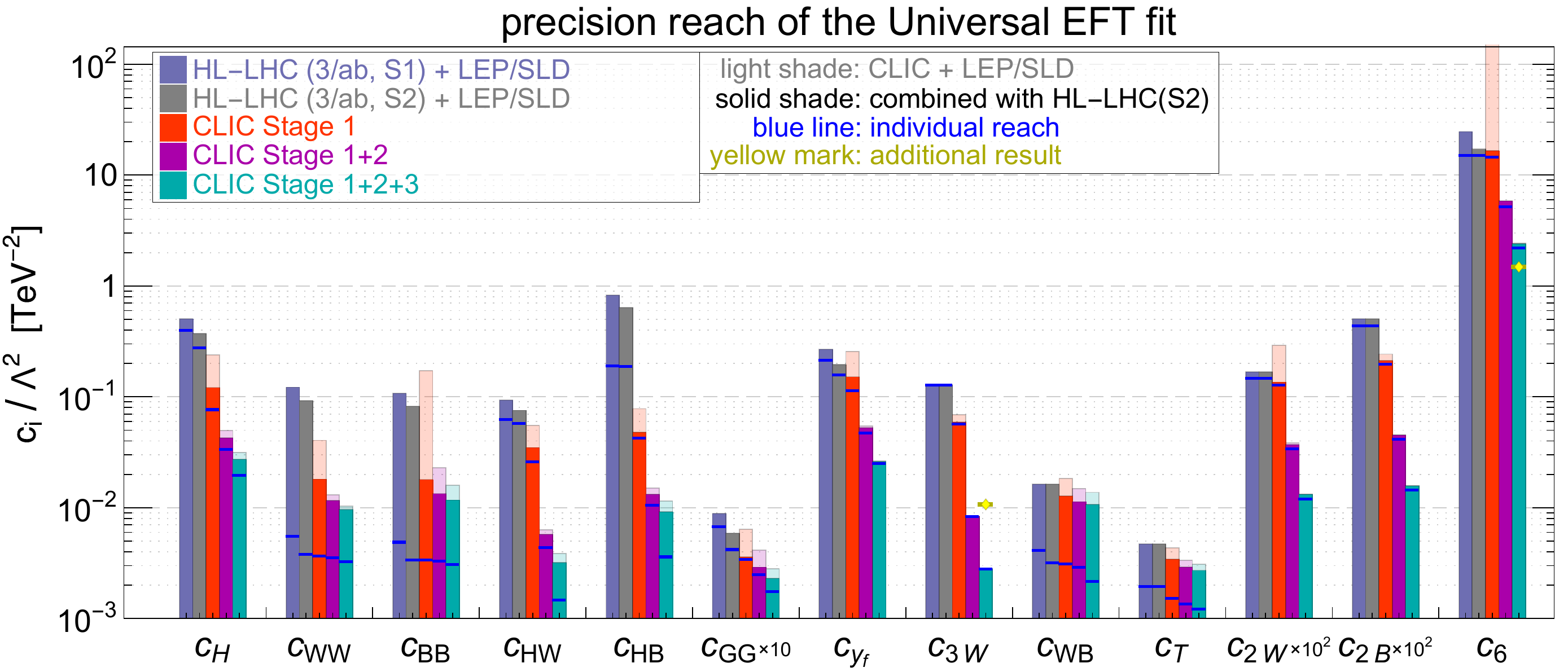}
\caption{Summary of the sensitivity to SM-EFT operators $c_i/\Lambda^2$ from a global analysis of CLIC's sensitivities to Higgs couplings, top-quark observables, $\PWp\PWm$ production, and two-fermion scattering processes $\epem\to f \overline{f}$, for the three CLIC energy stages~\cite{ESU18BSM}. Smaller values correspond to a higher scale probed. Preliminary projections for HL-LHC are shown for comparison, under two systematic uncertainty scenarios, S1 and S2. Blue markers correspond to single-operator sensitivies, and yellow markers correspond to results from dedicated individual analyses (for example, the Higgs self-coupling analysis).}
\label{fig:eft_limits_summary_esu}
\end{center}
\end{figure}
CLIC sensitivities to Higgs couplings, top-quark observables, $\PWp\PWm$ production,
and two-fermion scattering processes $\epem\to f \overline{f}$, where $f=\PQc, \PQb, \PQt, \Pe, \PGm, \PGt$, have all been combined
in a global fit using SM-EFT as introduced above in the context of top-quark pair-production. This approach is also well suited for a direct comparison to existing and other future options.
The CLIC sensitivity to the operator coefficients $c_i/\Lambda^2$ for
the operator basis defined in~\cite[Tab. 2]{ESU18BSM} is shown in \ref{fig:eft_limits_summary_esu}.
In this basis, the Higgs couplings to SM particles are mostly sensitive to $c_{\mathrm{BB}}$ for the $\PW$ and $\PZ$ bosons, $c_{\mathrm{GG}}$ for the gluons and $c_{\mathrm{y}_{f}}$ for the Yukawa coupling, while $c_{\PH}$ is a universal rescaling of all the Higgs couplings. They are probed to high accuracy in Higgs boson production and decay processes at CLIC.
$c_{\mathrm{WB}}$ and $c_{\mathrm{T}}$ contribute to Higgs couplings, but also to electroweak processes such as two-fermion production.
The parameter $c_6$ governs the Higgs self-coupling, and is tested through the direct double-Higgs boson production measurement at CLIC.
The operators corresponding to the parameters $c_{3\PW}$, $c_{2\PW}$, $c_{2\PB}$, $c_{\PH\PW}$ and $c_{\PH\PB}$ have effects that grow with energy.
Their measurements clearly benefit most from the high-energy stages.
$c_{3\PW}$ generates transverse anomalous triple gauge couplings which can be probed in di-boson production.
The parameters $c_{\PH\PW}$ and $c_{\PH\PB}$ are related to longitudinal anomalous triple gauge couplings as well as anomalous Higgs couplings, which influence both di-boson and Higgs boson production processes.
Finally, the kinetic terms of the electroweak gauge bosons, modified by $c_{2\PW}$ and $c_{2\PB}$, correspond to the $W$ and $Y$ parameters discussed above.
These results demonstrate that already the initial stage of CLIC is very
complementary to the HL-LHC for many of the operators~\cite[Sec.~2.9]{ESU18BSM}. 
The high-energy stages, which
are unique to CLIC among all proposed \epem colliders, are found to be
crucial for the precision programme.
Overall, CLIC probes the EFT operator coefficients much more precisely than is possible at the HL-LHC.

\section{New physics searches}\label{sec:DirectBSM}
The CLIC potential to explore concrete new physics scenarios, which address
several of the fundamental open questions of particle physics, is extensively 
documented in the literature and summarised in~\cite{Battaglia:2004mw,cdrvol2,cdrvol3,ESU18BSM}.
An incomplete selection is described in this section, in order to illustrate
the scope and reach of CLIC.

\paragraph{Direct discoveries of new particles}

CLIC can probe TeV-scale electroweak charged particles, or more generally particles that interact with the SM with electroweak-sized couplings, well above
the HL-LHC reach. Such new particles are expected because
many of the shortcomings of the Standard Model
are inherent to the electroweak sector of the theory. For example
particle dark matter candidates can hardly carry any SM charge other
than electroweak. Furthermore the electroweak sector may be able to
accommodate the violation of baryon number, C and CP necessary for
the generation of the baryon/anti-baryon asymmetry of the Universe, as well as to provide
a phase transition and the necessary boundaries between phases at
which to generate the asymmetry. All unsolved questions about the
origin of the masses and mixings of neutrinos and of the other fermions
of the SM are related to the weak interactions. In addition, the Naturalness
Problem is in essence a question about the peculiarity of weak interactions.
A complete exploration of TeV-scale electroweak particles is thus
a priority for particle physics. 

Any such new particle can be produced at CLIC with sizeable rate up
to the kinematic limit of 1.5\,TeV, and in some cases up to 3\,TeV via
single production mechanisms. Depending on the decay channels, different
detection strategies are possible. 

When new particles decay into standard final states featuring prompt
jets, leptons and photons they give rise to signatures that can be
distinguished relatively easily from backgrounds. Indeed, backgrounds
from SM processes usually have cross sections comparable to the signal,
as they are produced via the same electroweak interactions. This is
the key advantage of lepton colliders over hadron colliders that makes
CLIC outperform the HL-LHC. In most cases the signals can be isolated
so clearly that it is possible to measure with precision new particle
properties such as mass and spin, and even test concrete models of
BSM physics by checking some of their key predictions
on new particle properties. The CLIC potential to directly
explore new physics is extensively documented in the literature, and
in particular in~\cite{cdrvol2,cdrvol3}, often taking supersymmetric particles
as benchmarks. The CLIC direct reach for a variety of BSM scenarios,
ranging from non-minimal supersymmetry and extended Higgs sectors,
to Dark Matter, Baryogenesis and neutrino mass models, is presented
in~\cite{ESU18BSM}. Some of these are also documented
in the following.

\begin{wrapfigure}{R}{0.5\textwidth}
\includegraphics[width=\linewidth]{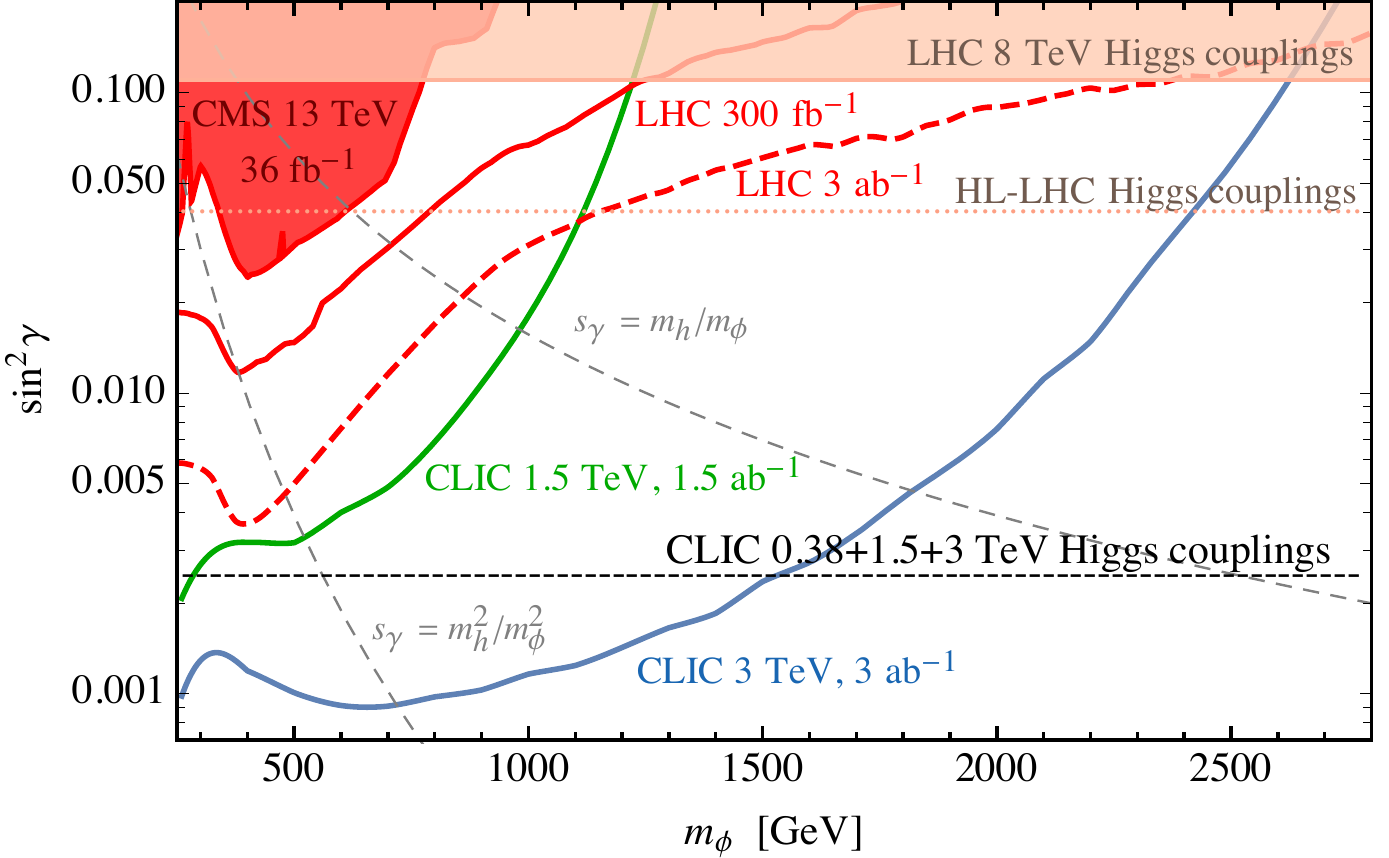}
\caption{\label{fig:Reach-for-Singlet}Reach for new scalar singlets in the mixing vs. mass plane~\cite{ESU18BSM}. Mixing above the horizontal dashed black line is excluded by single Higgs couplings measurements. Regions above the blue (green) lines are excluded by the direct search of $\HepParticle{S}{}{}\Xspace \to \Ph\Ph \to 4\PQb$. }
\vspace{-20pt}
\end{wrapfigure}

When the new particles give rise to non-standard signatures, e.g.
because they decay at a macroscopic distance in the detector volume,
it is still possible to isolate these signals thanks to the clean
environment typical of $\epem$ colliders. Relevant examples of this
kind of signature include Higgs boson rare decays to long-lived particles~\cite[Sec.~8]{ESU18BSM},
Higgsino Dark Matter, and the search for WIMP baryogenesis models.
These results are further discussed below.

\paragraph{Extended Higgs Sector}

Understanding the nature of the Higgs boson is one of the key elements
for a full understanding of the electroweak interactions and in particular
of the breaking of electroweak symmetry. A very important question
is whether the Higgs is the unique scalar particle at the weak scale, or
if instead an extended scalar sector exists. For this reason
a key target for future colliders is to investigate the existence of
additional Higgs bosons at the TeV scale, which may be the first important
step to unravelling the mystery of electroweak symmetry breaking and the
origin of the weak scale. 

A prototypical example of an extended Higgs sector is the extension of
the SM with a new scalar. A particularly challenging case is the one
in which the new scalar has no gauge interactions and interacts with
the SM only through the Higgs boson portal. This kind of scalar is usually
referred to as a ``singlet'' scalar and arises in concrete models such
as the Next-to-Minimal Supersymmetric Standard Model (NMSSM), non-minimal
Composite Higgs models as well as Twin Higgs models from ``neutral
naturalness'' solutions to the hierarchy problem of the weak scale.
In addition, such a new scalar may affect the Higgs potential
and alter the nature of the phase transition between broken and unbroken
electroweak symmetry in the early Universe, thus playing a role in
the generation of a net baryon number. 

A concrete study of direct production of a new scalar singlet at CLIC
is summarised in \ref{fig:Reach-for-Singlet}. CLIC sensitivity
to direct production of a new scalar singlet extends well beyond the
TeV mass scale, at which these new singlets are most motivated. For
a mixing between the singlet and the Higgs of $\sin^{2}\gamma<0.24\%$
the new singlet has to be heavier than 1.5\,TeV. Furthermore if a singlet
of \emph{any mass} has mixing $\sin^{2}\gamma>0.24\%$ it would result
in deviations in the single Higgs couplings to SM gauge bosons and
fermions in excess of 2 standard deviations for the expected statistical accuracy
of Higgs couplings determinations at CLIC.  These studies are discussed
in detail in~\cite[Sec.~4.2]{ESU18BSM}, where their implications
on concrete models are also worked out. It is found that in the case
of the NMSSM CLIC can exclude a new scalar lighter than 1.5\,TeV for
values of $\tan\beta \leq 4$, where the NMSSM is most motivated. For Twin
Higgs models direct searches bounds generically rule out new scalars
below 2\,TeV for values of the dynamical scale of the model $f<2\,\text{TeV}$,
where the model is most motivated. Furthermore, the study of Higgs
boson couplings yields bounds  $f>4.5\text{ TeV}$ if one assumes
that the mass of the scalar is equal or greater than $f$, as expected
for a composite scalar.
Results on several concrete models featuring extended scalar sectors
with multiple doublets and singlets fields, e.g. the 2-Higgs-Doublets
model (2HDM) are described in \cite[Sec.~4.3]{ESU18BSM}. CLIC can probe the existence of such new scalars both
by direct production and by the indirect effects they have
on the 125\,GeV Higgs boson couplings. The expected reach extends up
to and beyond 1\,TeV, improving dramatically on the HL-LHC searches.
 All in all CLIC is able to thoroughly test extended Higgs sectors
and rule out new scalars up to multi-TeV masses. Both direct and indirect
signatures can be successfully pursued yielding stringent bounds on
new scalar particles that improve by almost one order of magnitude
in mass scale on the HL-LHC projections.

\paragraph{Composite Higgs}

\begin{figure}[bpht]
\centering
\begin{subfigure}{0.45\textwidth}
  \includegraphics[width=\linewidth]{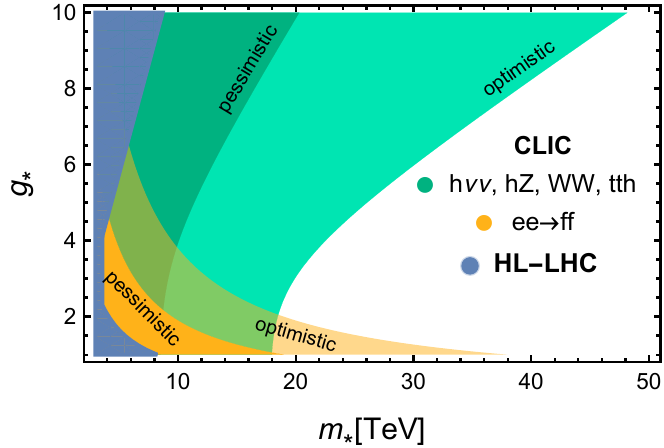}
\label{fig:ch}
\end{subfigure}
\hspace{30pt}
\begin{subfigure}{0.45\textwidth}
\centering \includegraphics[width=\linewidth]{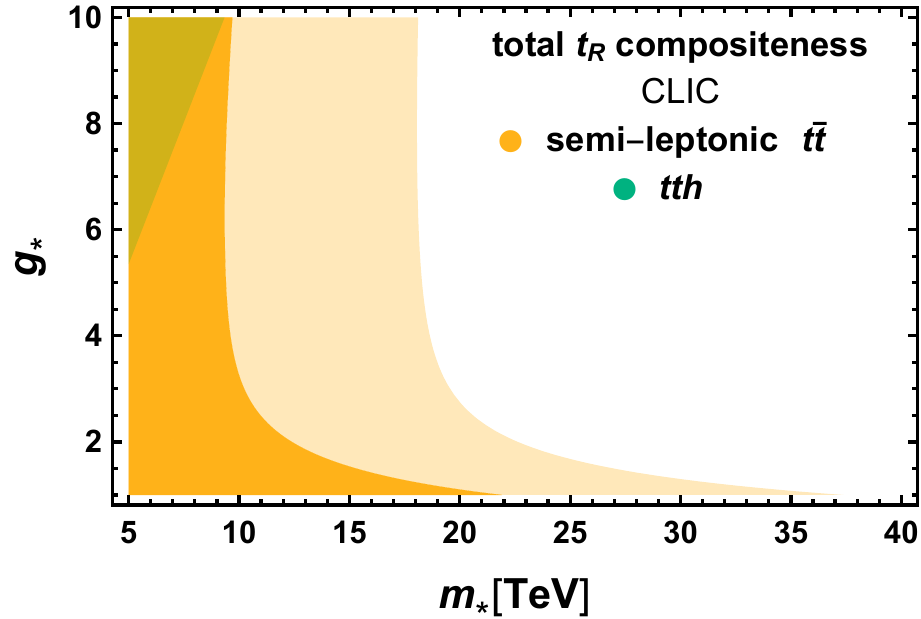}
\label{fig:ch-1}
\end{subfigure}
\vspace{-5pt}
\caption{$5\sigma$ discovery contours for (left) Higgs compositeness in the $(m_{*},g_{*})$
  plane, overlaid with the $2\sigma$ projected exclusions from HL-LHC and (right) top-quark compositeness~\cite{ESU18BSM}.}
\label{fig:compH}
\end{figure}

The Higgs boson is the only scalar particle that is predicted in the
SM to be exactly point-like. Therefore it is interesting to investigate
whether instead it is an extended composite object and, if it is, to determine
its geometric size $l_{\PH}$. Discovering the composite nature of the
Higgs would be a crucial step towards the understanding of the microscopic
origin of the 
electroweak symmetry breaking phenomenon. Higgs compositeness might
also solve or ameliorate the fine-tuning (or Naturalness) Problem
associated with the SM Higgs mass parameter. A composite Higgs would
manifest itself at CLIC through $d=6$ SM-EFT operators, suppressed
by two powers of the Higgs compositeness scale $m_{*}\sim1/l_{\PH}$.
The operator coefficients are enhanced or suppressed, relative to
the naive $1/m_{*}^{2}$ scaling, by positive or negative powers of
a parameter ``$g_{*}$'' representing the coupling strength of the
composite sector from which the Higgs emerges. These rules provide estimates
for the operator coefficients in the $(m_{*},g_{*})$ plane and allow
translation of the CLIC
sensitivity to the SM-EFT into the discovery reach on Higgs compositeness (\cite[Sec.~2.10]{ESU18BSM} and references therein),
as in \ref{fig:compH}(left). Given that the estimates only hold up to ``$c$'' coefficients of order one, 
the discovery contour is not sharply defined, but ranges from a
``pessimistic'' (for discovery) to an ``optimistic'' line. Those
are obtained by independently varying each $c$ coefficient in the
$[1/2,2]$ interval and selecting the configuration which is, respectively,
less or more favourable for discovery. 
The projected HL-LHC exclusion (as opposed to discovery, as in the
CLIC lines) reach~\cite[Sec.~2.10]{ESU18BSM} is also shown in the
figure, for unit $c$-coefficients. The dramatic improvement achieved
by CLIC at small and intermediate $g_{*}$ is due to the high-energy
stages that allow for a very precise determination of the $c_{HW}$,
$c_{HB}$, $c_{2W}$ and $c_{2B}$ SM-EFT coefficients. Single
Higgs boson couplings measurements instead provide the most stringent
constraints at large $g_{*}$. Within the same context, and in connection
with the Naturalness Problem, top-quark compositeness can also be
considered. It produces SM-EFT operators in the top sector that can
be probed by measuring the top Yukawa coupling and, very effectively,
by $\PQt\PAQt$ production at high-energy CLIC. The reach in
the ``total $t_{R}$ compositeness'' scenario is displayed in \ref{fig:compH}(right).
For further details, and for a similar result in the case of ``partial
top compositeness'', see~\cite[Sec.~2.10]{ESU18BSM} and
\cite[Sec.~10.2]{ClicTopPaper}.

\paragraph{Dark Matter}

\begin{wrapfigure}{R}{0.5\textwidth}
\vspace{-40pt}
\includegraphics[width=\linewidth]{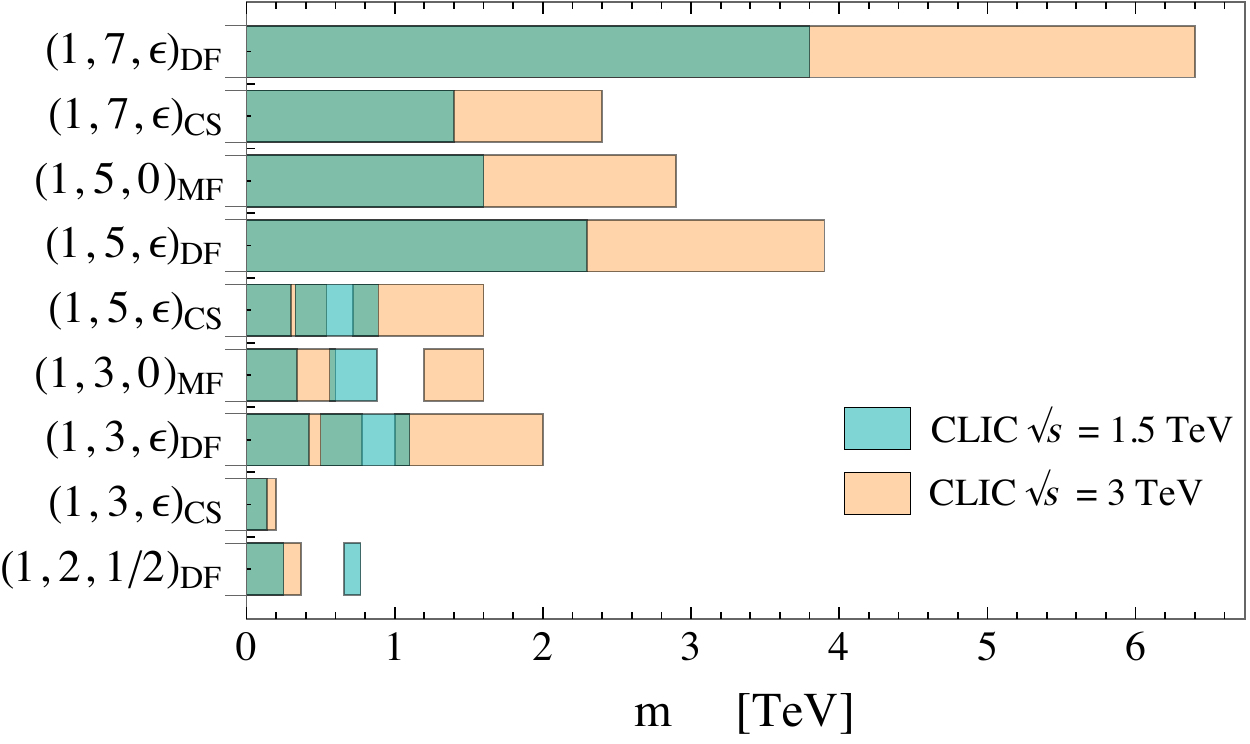}
\caption{\label{fig:reach-DM}Masses excluded at 95\%\,C.L. for new electroweak $n$-plet
states with hypercharge $Y$~\cite{ESU18BSM}. The exclusion for each state denoted
by (1,n,$Y$) for the higher-energy stages of CLIC is presented in the blue and
beige bars. Green regions would be excluded by data from both centre-of-mass energies.}
\vspace{5pt}
\end{wrapfigure}

The microscopic origin of the observed Dark Matter (DM) abundance
is one of the greatest mysteries of particle physics. Experimental
progress is possible at CLIC in well-motivated corners of the vast
landscape of viable DM models, as documented in~\cite[Sec.~5]{ESU18BSM}.
In particular if the DM relic abundance is produced by thermal freeze-out,
a compelling ``minimal'' possibility is that the DM annihilation
process proceeds through the familiar SM EW force, rather than through
a new not-yet observed interaction. This scenario requires TeV-scale
DM mass, and it realises the Weakly Interacting Massive Particle (WIMP)
miracle in its most appealing form. WIMP DM has been probed extensively
in direct detection experiments. However, other structurally
elusive candidates exist, such as the Higgsino doublet and the Wino triplet
or, more generally, the so-called ``Minimal DM'' particles that
can exist also in very large (up to the $7$-plet) representations
of the SM %
\mbox{%
SU$(2)$%
} group and can have multi-TeV masses. The latter candidates are a target for future colliders. CLIC
can probe them in several ways. First, model-independent
indirect searches for new EW states can be performed by studying their radiative effects
on the EW pair-production of SM particles. The 95\%\,C.L.
sensitivities for such searches at CLIC are reported in \ref{fig:reach-DM}. The sensitivity
reaches the thermal mass (i.e., the one which is needed in order to
produce the observed thermal abundance by standard thermal production) in the case of the Dirac fermion
triplet candidate $(1,3,\epsilon)_{{\rm {DF}}}$. Alternatively, one can
exploit the fact that the charged component of the Minimal DM multiplet
is long-lived, with a macroscopic decay length. Its distinctive signature
is thus a ``stub'' track, which can be long enough to be reconstructed at CLIC if
the particle is light enough to be sufficiently boosted.

CLIC can discover the thermal Higgsino at 1.1\,TeV with
this strategy. CLIC is also sensitive to DM models that fall outside
the Minimal DM paradigm, such as co-annihilation and Inert Doublet
models as documented in~\cite[Sec.~5]{ESU18BSM}. More generally, CLIC can effectively
probe DM models that have a sufficient mass-splitting to produce signals
featuring prompt jets, leptons, and photons plus missing momentum.

\paragraph{Baryogenesis and electroweak phase transition}

\begin{wrapfigure}{R}{0.4\textwidth}\vspace{-20pt}
\includegraphics[width=\linewidth]{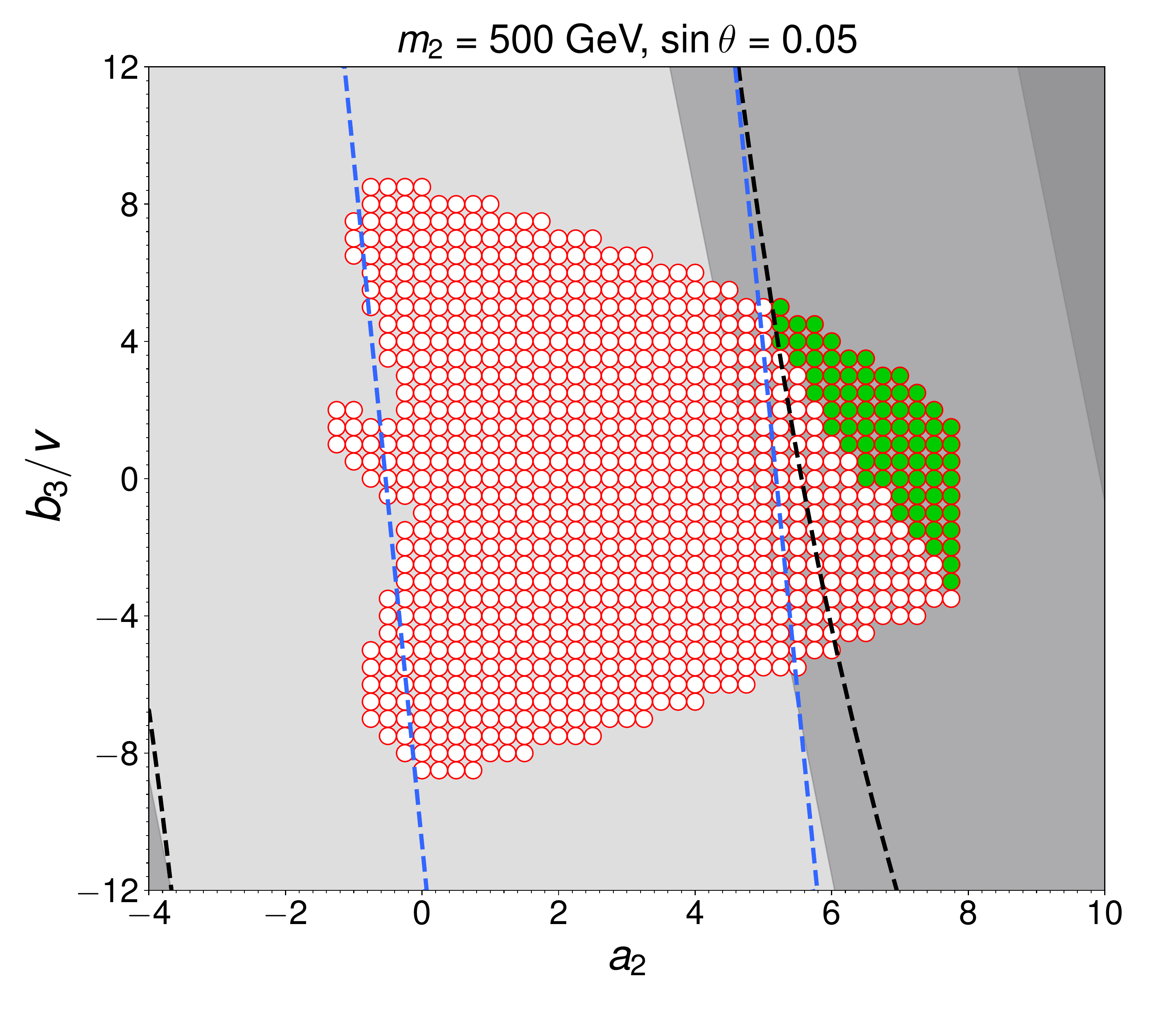}
\caption{\label{fig:ewbg-reach}Reach on the ``Higgs plus singlet'' model for electroweak
baryogenesis~\cite[Sec.~6.1]{ESU18BSM}. First order phase transition is allowed in the parameters
space marked by green points. Regions outside the blue (black) dashed
lines are excluded by CLIC direct search for $\HepParticle{S}{}{}\to \PH\PH$ (triple Higgs
coupling mea\-surement).}
\vspace{-10pt}
\end{wrapfigure}

The mechanism responsible for the origin of the asymmetry between
baryons and anti-baryons in the Universe is currently unknown, and
it might be discovered at CLIC if it is related to TeV-scale physics,
as in the electroweak baryogenesis scenario. This mechanism
requires, among other things, a considerable modification of the SM
thermal Higgs potential that should give rise to an EW phase transition
of strong first order, unlike the smooth crossover that is predicted
by the SM. This is achieved by postulating the existence of new scalar
particles coupled with the Higgs, which can be exhaustively probed
at CLIC by precise measurements of the Higgs trilinear coupling and
of single Higgs couplings, and by direct searches. This is illustrated
in \ref{fig:ewbg-reach}, where we report the 95\%\,C.L.
CLIC reach on the ``Higgs plus singlet'' model~\cite[Sec.~6.1]{ESU18BSM}. This is
a suitable illustrative benchmark because it contains the minimal
amount of new physics (i.e. a scalar singlet $\HepParticle{S}{}{}$) that is needed
to achieve a strong first-order phase transition. The singlet is described
by the most general renormalisable Lagrangian and it couples to the
SM only through Higgs-portal interactions. The figure displays a slice
of the parameter space of the model for singlet mass $m_{2}=500$~GeV
and singlet mixing $\sin\theta=0.05$ with the SM Higgs boson. The remaining
parameters $a_{2}$ and $b_{3}$ are respectively the $|H|^{2}S^{2}$
quadratic portal coupling and the $S^{3}$ trilinear vertex. The allowed
points in the plane are marked as red circles, and those for which the
EW phase transition is strong enough are filled in green. All the
green points can be probed both by the trilinear Higgs coupling measurement
(black dashed) and by single $S$ production decaying to $\PH\PH \to 4\PQb$
final state (blue dashed). For this particular choice of mixing, the entire plane
is probed by single Higgs coupling measurements (grey region). Note
however that the effectiveness of single-Higgs couplings (which would be dominant even for $\sin\theta = 0$ due to loop-induced contributions) stems from
the fact that this model predicts sharp correlations between
the modification of several Higgs vertices, but these correlations
might be relaxed in other models. The CLIC capability to perform
multiple competitive probes of the scenario instead allows
robust conclusions to be drawn~\cite[Sec.~4.2]{ESU18BSM}.

CLIC can also probe TeV-scale Baryogenesis models of radically different
nature. In particular the so-called ``WIMP baryogenesis''
scenario can be considered, where the baryon asymmetry is generated via the baryon number
violating decays of TeV-scale long-lived particles. The favourable
experimental conditions of CLIC allow unexplored regions
of the mass-lifetime parameter space of this model to be probed~\cite[Sec.~6.2]{ESU18BSM}. In particular,
long decay lengths, which are necessary to generate
necessary out-of-equilibrium decays in the early Universe, can be explored.

\paragraph{Hidden Sector}

New physics could manifest itself with light new particles, which
have not yet been seen on account of their tiny couplings with SM particles.
CLIC can make progress~\cite[Sec.~8]{ESU18BSM} on the
experimental exploration of this scenario in unique corners of its
vast parameter space, including particularly well-motivated ones.
For example, the clean environment and the absence of a trigger allows
CLIC to improve significantly over the HL-LHC in the search for Higgs
or Higgs-like bosons decaying to long-lived particles. This will allow
CLIC to probe the Fraternal Twin Higgs and Folded Supersymmetry solutions
of the Naturalness Problem. CLIC can also search for heavy Axion-Like
Particles, in a mass-range that is manifestly outside the reach of
dedicated low-energy experiments. 

\paragraph{Neutrino Mass}

Neutrino masses cannot be explained in the Standard Model and they
require new physics either in the form of a chiral partner of the
left-handed neutrinos, i.e. as a heavy right-handed neutrino, or
in the form of new contact interactions between leptons and the Higgs
boson. These interactions happen to be non-renormalisable. Hence they
require some new physics at higher mass scales to originate them.
CLIC has sensitivity to a large set of models in which lepton number
is an almost approximate symmetry, which makes it very natural to have
small neutrino masses in the form of a Majorana mass. Detailed results
can be found in~\cite[Sec.~7]{ESU18BSM}; here, we give
some representative results. For example, in the inverse-see-saw model
it is possible to have large Yukawa couplings between the Higgs boson
and the right-handed neutrinos, which CLIC can exclude up to 10\,TeV
mass when the Yukawa coupling is of order one. Even greater reach around
tens of TeV is expected for models featuring a doubly charged scalar
lepton for Yukawa coupling of order one. Furthermore, CLIC can easily
exclude the presence of electroweak charged scalars and fermions,
such as the mediators of type-2 and type-3 see-saw models, with masses
below 1.5\,TeV over the entire parameter space of the models. In particular,
for type-2 see-saw CLIC is able to probe the model for any value
of the Vacuum Expectation Value (VEV) of the triplet neutral scalar, dramatically improving on
the situation of the HL-LHC which is hardly sensitive when the VEV is greater than 100\,keV.

\paragraph{Flavour violation}

The high available energy and the clean experimental conditions make
CLIC extremely sensitive to $4$-fermion SM-EFT operators that produce
Flavour-Changing Neutral Current (FCNC) effects in the lepton and in the quark sector. Operators with the structure
$(\Pe\Pe)(\PGt\Pe)$ and $(\Pe\Pe)(\PQt\PQq)$ induce, respectively, $\PGt$ plus
$\Pe$ and top plus light jet production processes that are extremely
rare in the SM, at a rate that grows with the fourth power of the
centre-of-mass energy. By searching for these final states, CLIC is
sensitive to operator scales in excess of $40$\,TeV~\cite[Sec.~3.1]{ESU18BSM}. A comparable
sensitivity on $\PGt$ FCNC operators might be achieved at Belle 2
and at the HL-LHC from $\PGt\to\Pe\Pe\Pe$ decays. Top sector operators
instead have little chance to be probed at the same level in top decays,
where they would produce a tiny exotic branching ratio below around
$10^{-12}$. Current projections~\cite{MartinCamalich:2650175} show that more than one
order of magnitude improvement in the operator scale
is possible at CLIC with respect to the HL-LHC.

\paragraph{Summary of the CLIC discovery reach}

A brief summary of the CLIC discovery reach for the scenarios discussed here and others is given in \ref{tab:discovery_reach}. Where possible, a comparison with HL-LHC projections is given.

\begin{table}[ht]\centering
\caption{Indicative CLIC reach for new physics. Sensitivities are given for the full CLIC programme covering the three centre-of-mass energy stages. All limits are at 95\%\,C.L. unless stated otherwise.\label{tab:discovery_reach} Details on many of these examples are given in \cite{ESU18BSM}.}
\begin{adjustbox}{width=\linewidth}
\begin{tabular}{lll}\toprule
\textbf{Process} & \textbf{HL-LHC} & \textbf{CLIC} \\
   \hline
Higgs mixing with heavy singlet & $\sin^2\gamma<\SI{4}{\percent}$ & $\sin^2\gamma<\SI{0.24}{\percent}$ \\
Higgs self-coupling $\Delta\lambda$      & $\sim 50\%$ at \SI{68}{\percent} C.L. & $[\SI{-7}{\percent},\SI{+11}{\percent}]$ at \SI{68}{\percent} C.L.\\ 
\br{\PH\to \mathrm{inv.}} \footnotesize{(model-independent)}                                   &  & $<\SI{0.69}{\percent}$ at \SI{90}{\percent} C.L. \\
\hline
Higgs compositeness scale $m_*$  &  $m_*>\SI{3}{\TeV}$ & Discovery up to $m_*$ = $\SI{10}{\TeV}$ \\
                                 &  \;\;\;\;\; ($>\SI{7}{\TeV}$ for $g_*\simeq 8$) &  \;\;\;\;\; ($\SI{40}{\TeV}$ for $g_*\simeq 8$) \\
\hline
Top compositeness scale $m_*$    &  & Discovery up to $m_*$ = $\SI{8}{\TeV}$ \\
                                 &  &  \;\;\;\;\; ($\SI{20}{\TeV}$ for small coupling $g_*$) \\
\hline
Higgsino mass {\footnotesize{(disappearing track search)}}  & $>\SI{250}{\GeV}$ & $>\SI{1.2}{\TeV}$ \\ 
Slepton mass                                                     &  & Discovery up to $\sim 1.5$\,\TeV \\ 
RPV wino mass \footnotesize{($c\tau = \SI{300}{\meter}$)}                        & $>\SI{550}{\GeV}$ & $>\SI{1.5}{\TeV}$  \\ 
\hline
$\PZ'$ mass \footnotesize{({SM couplings})}            & Discovery up to 7\,TeV & Discovery up to 20\,TeV \\ 
\hline
NMSSM scalar singlet mass     &  $>\SI{650}{\GeV}$ ($\tan\beta \leq 4$)  & $>\SI{1.5}{\TeV}$ ($\tan\beta \leq 4$) \\ 
Twin Higgs scalar singlet mass                                    & $m_{\sigma} = f > 1$\,TeV & $m_{\sigma} = f > 4.5$\,TeV \\ 
\hline
Relaxion mass \footnotesize{(for vanishing mixing)}                            & $<\SI{24}{\GeV}$ & $<\SI{12}{\GeV}$  \\ 
Relaxion mixing angle \footnotesize{($m_{\phi} < m_{\PH} / 2$)}                             &  & $\sin^2\theta\leq \SI{2.3}{\percent}$ \\
\hline
Neutrino Type-2 see-saw triplet                                  &  & $>\SI{1.5}{\TeV}$ ({\footnotesize{for any triplet VEV}}) \\
                                                                 &  & $>\SI{10}{\TeV}$ ({\footnotesize{for triplet Yukawa coupling $\simeq 0.1$}}) \\
\hline
Inverse see-saw RH neutrino                                     &  & $>\SI{10}{\TeV}$ ({\footnotesize{for Yukawa coupling $\simeq 1$}}) \\
\hline
Scale $V_{LL}^{-1/2}$ for LFV $(\bar{\Pe}\Pe)(\bar{\Pe}\PGt)$     &  & $>\SI{42}{\TeV}$ \\ 
\bottomrule
\end{tabular}
\end{adjustbox}
\end{table}

\section{Conclusions}

CLIC offers the unique combination of high collision energies and the clean environment of electron--positron collisions.
This enables the guaranteed physics programme of SM-parameter measurements with unprecedented precision, ranging from the top-quark mass and other top-quark properties to the Higgs couplings, including the Higgs self-coupling.
In addition, parameters describing deviations from the SM in a model-independent way can be constrained at CLIC with a precision well exceeding that of the HL-LHC.
This includes the electroweak parameters $W$ and $Y$ measured in two-fermion production, and EFT coefficients measured in Higgs and electroweak processes.
The staged running in three energy stages at $\sqrt{s} = 380$\,GeV, 1.5\,TeV, and 3\,TeV is important not only to get optimal access to the various processes, but also to probe energy-dependent operators to a level of precision that exceeds that of the HL-LHC by more than one order of magnitude.

Furthermore, CLIC offers a rich potential for extensive exploration of the terascale in the form of direct and indirect searches of BSM effects. Direct searches are often possible up to the kinematic limit for particles with electroweak-sized coupling strength and detectable decay products. New physics effects, for example from scalars in an extended Higgs sector or from a composite Higgs sector, can be found directly or, beyond the kinematic reach, through effects of their mixing with known particles measured in Higgs boson production at CLIC. Long-lived charged particles such as the charged component of minimal dark matter multiplets can give rise to disappearing tracks, for which the clean environment and the detector layout are well suited. The measurement of double Higgs boson production will constrain models of electroweak baryogenesis. Other signatures include the measurement of soft decay products of new particles, e.g. from hidden sectors, which benefits from triggerless running and the clean environment. Additionally, high statistics of top quarks and Higgs bosons allow the search for rare decays indicating for example flavour violation effects. Direct and indirect searches for TeV-scale mediators of neutrino mass generation can provide further insights into the physics of flavour.

The current situation of particle physics is that experiments up to now, including those running at the LHC, could not provide answers to many of the open questions on fundamental interactions. The quest for BSM physics is even more pressing now than in the past. On the other hand, there are no robust hints
of new physics at a nearby energy threshold. In this context, an innovative and ambitious project like CLIC with a far-reaching programme of direct and indirect BSM searches to extend the borders of our knowledge into unexplored territories is highly desirable. CLIC combines this with a guaranteed outcome of precision measurements of SM processes and emerges in this context as a particularly appealing option for the future of high energy physics.

\newpage
\printbibliography[title=References]

\clearpage
\appendix
\section{Addendum}
\subsection{Community}\label{addendum:community}
The CLIC accelerator and CLIC Detector and Physics collaborations together comprise around 400 participants from approximately 75 institutes worldwide~\cite{clic-study}.
Additional contributions are made from beyond the collaborations, in particularly from the
phenomenology community participating in the wider CLIC Physics Potential Working Group.
This document is based on the work reported in the Compact Linear Collider (CLIC) 2018 Summary Report~\cite{ESU18Summary} and the CLIC Potential for New Physics Report~\cite{ESU18BSM}.  The authors of these two reports are listed below.

\FloatBarrier

\vspace{2ex}\centering \textbf{{The CLIC Potential for New Physics}}
{\footnotesize
\begin{multicols}{2}

\begin{center}
\textbf{Editors:}\\
J.~de~Blas $^{1,2}$, R.~Franceschini $^{3,4}$,  F.~Riva $^5$,  P.~Roloff $^6$,  U.~Schnoor $^6$,  M.~Spannowsky $^7$,  J.~D.~Wells $^{8}$,  A.~Wulzer $^{1,6,9}$  and J.~Zupan $^{10}$ \\
\vspace{10pt}
\flushleft
$^1$Dipartimento di Fisica e Astronomia ``Galileo Galilei'', Universit\`a di Padova, Padova, Italy\\ 
$^2$INFN, Sezione di Padova, Padova, Italy\\
$^3$Dipartimento di Matematica e Fisica, Universit\`a degli Studi Roma Tre, Rome, Italy\\
$^4$INFN, Sezione di Roma Tre, Rome, Italy\\
$^5$D\'epartment de Physique Th\'eorique, Universit\'e de Gen\`eve, Gen\`eve, Switzerland\\
$^6$CERN, Geneva, Switzerland\\
$^7$Institute for Particle Physics Phenomenology, Department of Physics, Durham University, UK\\
$^8$Leinweber Center for Theoretical Physics, Physics Department, University of Michigan, USA\\
$^9$Theoretical Particle Physics Laboratory (LPTP), EPFL, Lausanne, Switzerland\\
$^{10}$Department of Physics, University of Cincinnati, Cincinnati, Ohio, USA

\end{center}
\noindent{\textbf{Contributors:}}\\
{
\mbox{S.~Alipour-Fard $^{1}$,} 
\mbox{W.~Altmannshofer $^{2}$,} 
\mbox{A.~Azatov $^{3,4}$,} 
\mbox{D.~Azevedo $^{5,6}$,} 
\mbox{J.~Baglio $^{7}$,} 
\mbox{M.~Bauer $^{8}$,} 
\mbox{F.~Bishara $^{9,10}$,} 
\mbox{J.-J.~Blaising $^{11}$,} 
\mbox{S.~Brass $^{12}$,} 
\mbox{D.~Buttazzo $^{13}$,} 
\mbox{Z.~Chacko $^{14,15}$,} 
\mbox{N.~Craig $^{1}$,} 
\mbox{Y.~Cui $^{16}$,} 
\mbox{D.~Dercks $^{9,17}$,} 
\mbox{P.~S.~B.~Dev $^{18}$,}
\mbox{L.~Di~Luzio $^{8,13,19}$,} 
\mbox{S.~Di~Vita $^{20}$,} 
\mbox{G.~Durieux $^{9,21}$,} 
\mbox{J.~Fan $^{22}$,} 
\mbox{P.~Ferreira $^{5,23}$,} 
\mbox{C.~Frugiuele $^{24}$,} 
\mbox{E.~Fuchs $^{24}$,} 
\mbox{I.~Garc\'ia $^{25,26}$,} 
\mbox{M.~Ghezzi $^{7,27}$,} 
\mbox{A.~Greljo $^{26}$,} 
\mbox{R.~Gr\"ober $^{8,28}$,} 
\mbox{C.~Grojean $^{9,28}$,}
\mbox{J.~Gu $^{29}$,} 
\mbox{R.~Hunter $^{30}$,}  
\mbox{A.~Joglekar $^{16}$,} 
\mbox{J.~Kalinowski  $^{31}$,} 
\mbox{W.~Kilian $^{12}$,} 
\mbox{C.~Kilic $^{32}$,} 
\mbox{W.~Kotlarski $^{33}$,} 
\mbox{M.~Kucharczyk $^{34}$,} 
\mbox{E.~Leogrande  $^{26}$,} 
\mbox{L.~Linssen $^{26}$,} 
\mbox{D.~Liu $^{35}$,} 
\mbox{Z.~Liu $^{14,15}$,} 
\mbox{D.~M.~Lombardo $^{36}$,} 
\mbox{I.~Low $^{35,37}$,} 
\mbox{O.~Matsedonskyi $^{24}$,} 
\mbox{D.~Marzocca  $^{4}$,} 
\mbox{K.~Mimasu $^{38}$,} 
\mbox{A.~Mitov $^{39}$,} 
\mbox{M.~Mitra $^{40}$,} 
\mbox{R.~N.~Mohapatra $^{14}$,} 
\mbox{G.~Moortgat-Pick $^{9,17}$,} 
\mbox{M.~M\"uhlleitner $^{41}$,} 
\mbox{S.~Najjari $^{42}$,} 
\mbox{M.~Nardecchia  $^{4,26}$,} 
\mbox{M.~Neubert $^{29,43}$,}
\mbox{J.~M.~No $^{44}$,} 
\mbox{G.~Panico $^{9,45,46,47}$,} 
\mbox{L.~Panizzi $^{48,49}$,} 
\mbox{A.~Paul $^{9,28}$,} 
\mbox{M.~Perell\'o $^{25}$,} 
\mbox{G.~Perez $^{24}$,} 
\mbox{A.~D.~Plascencia $^{8}$,} 
\mbox{G.~M.~Pruna $^{50}$,} 
\mbox{D.~Redigolo $^{24,51,52}$,} 
\mbox{M.~Reece $^{53}$,} 
\mbox{J.~Reuter $^{9}$,} 
\mbox{M.~Riembau $^{36}$,} 
\mbox{T.~Robens $^{54,55}$,} 
\mbox{A.~Robson $^{26,56}$,}
\mbox{K.~Rolbiecki $^{31}$,} 
\mbox{A.~Sailer $^{26}$,} 
\mbox{K.~Sakurai $^{31}$,} 
\mbox{F.~Sala $^{9}$,} 
\mbox{R.~Santos $^{5,23}$,} 
\mbox{M.~Schlaffer $^{24}$,} 
\mbox{S.~Y.~Shim $^{57}$,} 
\mbox{B.~Shuve $^{16,58}$,} 
\mbox{R.~Simoniello  $^{26,59}$,} 
\mbox{D.~Soko\l owska  $^{31,60}$,} 
\mbox{R.~Str\"{o}m  $^{26}$,} 
\mbox{T.~M.~P.~Tait $^{61}$,} 
\mbox{A.~Tesi $^{46}$,} 
\mbox{A.~Thamm $^{26}$,} 
\mbox{N.~van~der~Kolk $^{62}$,} 
\mbox{T.~Vantalon $^{9}$,} 
\mbox{C.~B.~Verhaaren $^{63}$,} 
\mbox{M.~Vos $^{25}$,} 
\mbox{N.~Watson  $^{64}$,} 
\mbox{C.~Weiland $^{8,65}$,} 
\mbox{A.~Winter $^{64}$,} 
\mbox{J.~Wittbrodt $^{9}$,} 
\mbox{T.~Wojton $^{34}$,} 
\mbox{B.~Xu  $^{39}$,} 
\mbox{Z.~Yin $^{37}$,} 
\mbox{A.~F.~\.Zarnecki $^{31}$,} 
\mbox{C.~Zhang  $^{66}$,}
\mbox{Y.~Zhang $^{18}$} 
}\\[20pt]
\flushleft
$^1$Department of Physics, University of California, Santa Barbara, CA, USA\\
$^2$Santa Cruz Institute for Particle Physics, Santa Cruz, CA, USA\\
$^3$SISSA, Trieste, Italy\\
$^4$INFN, Sezione di Trieste, Trieste, Italy \\
$^5$Centro de F\'isica Te\'orica e Computacional, Universidade de Lisboa, Campo Grande, Lisboa, Portugal\\
$^6$LIP, Departamento de F\'isica, Universidade do Minho, Braga, Portugal\\
$^7$Institute of Theoretical Physics, Eberhard Karls Universit\"{a}t T\"{u}bingen, Germany\\
$^8$Institute for Particle Physics Phenomenology, Department of Physics, Durham University, UK\\
$^9$DESY, Hamburg, Germany\\
$^{10}$Rudolf Peierls Centre for Theoretical Physics, University of Oxford, UK\\
$^{11}$Laboratoire d'Annecy-le-Vieux de Physique des Particules, France\\
$^{12}$Department of Physics, University of Siegen, Siegen, Germany\\
$^{13}$INFN, Sezione di Pisa, Pisa, Italy\\
$^{14}$Maryland Center for Fundamental Physics, Department of Physics, University of Maryland, USA \\
$^{15}$Theoretical Physics Department, Fermi National Accelerator Laboratory, Batavia, IL, USA  \\
$^{16}$Department of Physics and Astronomy, University of California, Riverside, CA, USA \\
$^{17}$II. Institute f. Theo. Physics, University of Hamburg, Hamburg, Germany \\
$^{18}$Dept.~\!of Physics and McDonnell Center for the Space Sciences, Washington University, St.~\!Louis, USA\\
$^{19}$Dipartimento di Fisica dell'Universit\`a di Pisa, Italy \\
$^{20}$INFN, Sezione di Milano, Milano, Italy \\
$^{21}$Physics Department, Technion --- Israel Institute of Technology, Haifa, Israel \\
$^{22}$Physics Department, Brown University, Providence, RI, USA \\
$^{23}$Instituto Superior de Engenharia de Lisboa, Instituto Polit\'ecnico de Lisboa, Lisboa, Portugal \\
$^{24}$Weizmann Institute of Science, Rehovot, Israel \\
$^{25}$IFIC (UV/CSIC) Valencia, Spain \\
$^{26}$CERN, Geneva, Switzerland \\
$^{27}$Paul Scherrer Institut, Villigen PSI, Switzerland \\
$^{28}$Institut f\"ur Physik, Humboldt-Universit\"at zu Berlin, Berlin, Germany \\
$^{29}$PRISMA Cluster of Excellence, Institut f\"ur Physik, Johannes Gutenberg-Universit\"at, Mainz, Germany \\
$^{30}$Department of Physics, University of Warwick, Coventry, UK \\
$^{31}$Faculty of Physics, University of Warsaw, Warsaw, Poland \\
$^{32}$Theory Group, Department of Physics, The University of Texas at Austin, Austin, TX, USA \\
$^{33}$Institut f\"ur Kern- und Teilchenphysik, TU Dresden, Dresden, Germany \\
$^{34}$The Henryk Niewodniczanski Institute of Nuclear Physics, Polish Academy of Sciences, Poland \\
$^{35}$High Energy Physics Division, Argonne National Laboratory, Argonne, IL, USA \\
$^{36}$D\'epartment de Physique Th\'eorique, Universit\'e de Gen\`eve, Gen\`eve, Switzerland \\
$^{37}$Department of Physics and Astronomy, Northwestern University, Evanston, IL, USA \\
$^{38}$Centre for Cosmology, Particle Physics and Phenomenology, UCLouvain, Belgium \\
$^{39}$Cavendish Laboratory, University of Cambridge, Cambridge, UK \\
$^{40}$Institute of Physics (IOP), Sachivalaya Marg, Bhubaneswar, India \\
$^{41}$Institute for Theoretical Physics, Karlsruhe Institute of Technology, Karlsruhe, Germany \\
$^{42}$Theoretische Natuurkunde and IIHE/ELEM, Vrije Universiteit Brussel, Brussels, Belgium \\
$^{43}$Department of Physics and LEPP, Cornell University, Ithaca, NY, USA \\ 
$^{44}$Departamento de F\'isica Te\'orica and Instituto de F\'isica Te\'orica, IFT-UAM/CSIC, Madrid, Spain \\
$^{45}$Dipartimento di Fisica e Astronomia Universit\`a di Firenze, Sesto Fiorentino, Italy \\
$^{46}$INFN, Sezione di Firenze, Italy \\
$^{47}$IFAE and BIST, Universitat Aut\`onoma de Barcelona, Bellaterra, Barcelona, Spain \\
$^{48}$Uppsala University, Uppsala, Sweden \\
$^{49}$School of Physics and Astronomy, University of Southampton, Southampton, UK \\
$^{50}$INFN, Laboratori Nazionali di Frascati, Frascati, Italy \\
$^{51}$Raymond and Beverly Sackler School of Physics and Astronomy, Tel-Aviv University, Tel-Aviv, Israel \\
$^{52}$Institute for Advanced Study, Princeton, NJ, USA \\
$^{53}$Department of Physics, Harvard University, Cambridge, MA, USA \\ 
$^{54}$MTA-DE Particle Physics Research Group, University of Debrecen, Debrecen, Hungary \\
$^{55}$Theoretical Physics Division, Rudjer Boskovic Institute, Zagreb, Croatia \\
$^{56}$University of Glasgow, Glasgow, UK \\
$^{57}$Department of Physics, Konkuk University, Seoul, Korea \\
$^{58}$Harvey Mudd College, 301 Platt Blvd, Claremont, CA, USA \\
$^{59}$Johannes Gutenberg-Universit\"at Mainz, Mainz, Germany \\
$^{60}$International Institute of Physics, Universidade Federal do Rio Grande do Norte, Brazil \\
$^{61}$Department of Physics and Astronomy, University of California, Irvine, CA, USA \\
$^{62}$Max-Planck-Institut f\"ur Physik, Munich, Germany \\
$^{63}$Center for Quantum Mathematics and Physics, University of California, Davis, CA, USA \\
$^{64}$University of Birmingham, Birmingham, UK \\
$^{65}$PITT PACC, Department of Physics and Astronomy, University of Pittsburgh, PA, USA \\
$^{66}$Institute of High Energy Physics, Chinese Academy of Sciences, Beijing, China 
\end{multicols}
}
\vspace{2ex}\centering \textbf{{The Compact Linear \boldmath{$\epem$} Collider (CLIC) -- 2018 Summary Report}}
{\footnotesize
\begin{multicols}{2}
\begin{center}{
T.K.~Charles, 
P.J.~Giansiracusa,
T.G.~Lucas,
R.P.~Rassool,
M.~Volpi$^{1}$\\
}\textbf{University of Melbourne, Melbourne, Australia}
\end{center}
\vspace{-0.5cm}

\begin{center}{
C.~Balazs\\ 
}\textbf{Monash University, Melbourne, Australia}
\end{center}
\vspace{-0.5cm}

\begin{center}{
K.~Afanaciev, 
V.~Makarenko\\
}\textbf{Belarusian State University, Minsk, Belarus}
\end{center}
\vspace{-0.5cm}

\begin{center}{
A.~Patapenka, 
I.~Zhuk\\
}\textbf{Joint Institute for Power and Nuclear Research - Sosny, Minsk, Belarus}
\end{center}
\vspace{-0.5cm}

\begin{center}{
C.~Collette\\
}\textbf{Universit\'e libre de Bruxelles, Brussels, Belgium}
\end{center}
\vspace{-0.5cm}

\begin{center}{
M.J.~Boland\\   
}\textbf{University of Saskatchewan, Saskatoon, Canada}
\end{center}
\vspace{-0.5cm}

\begin{center}{
A.C.~Abusleme~Hoffman,
M.A.~Diaz,
F.~Garay\\
}\textbf{Pontificia Universidad Cat\'{o}lica de Chile, Santiago, Chile}
\end{center}
\vspace{-0.5cm}

\begin{center}{
Y.~Chi,
X.~He,
G.~Pei,
S.~Pei,
G.~Shu,
X.~Wang,
J.~Zhang,
F.~Zhao,
Z.~Zhou\\
}\textbf{Institute of High Energy Physics, Beijing, China}
\end{center}
\vspace{-0.5cm}

\begin{center}{
H.~Chen,
Y.~Gao,
W.~Huang,
Y.P.~Kuang,
B.~Li,
Y.~Li,
X.~Meng, 
J.~Shao,
J.~Shi,
C.~Tang,
P.~Wang, 
X.~Wu,
H.~Zha\\ 
}\textbf{Tsinghua University, Beijing, China}
\end{center}
\vspace{-0.5cm}

\begin{center}{
L.~Ma,
Y.~Han\\
}\textbf{Shandong University, Jinan, China}
\end{center}
\vspace{-0.5cm}

\begin{center}{
W.~Fang,
Q.~Gu, 
D.~Huang, 
X.~Huang, 
J.~Tan, 
Z.~Wang, 
Z.~Zhao\\
}\textbf{Shanghai Institute of Applied Physics, Chinese Academy of Sciences, Shanghai, China}
\end{center}
\vspace{-0.5cm}

\begin{center}{
U.I.~Uggerh{\o}j,
T.N.~Wistisen\\
}\textbf{Aarhus University, Aarhus, Denmark}
\end{center}
\vspace{-0.5cm}

\begin{center}{
A.~Aabloo,
R.~Aare, 
K.~Kuppart,
S.~Vigonski,
V.~Zadin\\
}\textbf{University of Tartu, Tartu, Estonia}
\end{center}
\vspace{-0.5cm}

\begin{center}{
M.~Aicheler,
E.~Baibuz,
E.~Br\"{u}cken,
F.~Djurabekova$^{2}$,
P.~Eerola$^{2}$,
F.~Garcia,
E.~Haeggstr\"{o}m$^{2}$,
K.~Huitu$^{2}$,
V.~Jansson$^{2}$,
I.~Kassamakov$^{2}$,
J.~Kimari$^{2}$,  
A.~Kyritsakis,
S.~Lehti,
A.~ Meril\"{a}inen$^{2}$,
R.~Montonen$^{2}$,
K.~Nordlund$^{2}$,
K.~\"{O}sterberg$^{2}$,
A.~Saressalo,  
J.~V\"{a}in\"{o}l\"{a},
M.~Veske\\
}\textbf{Helsinki Institute of Physics, University of Helsinki, Helsinki, Finland}
\end{center}
\vspace{-0.5cm}

\begin{center}{
W.~Farabolini,
A.~Mollard,
F.~Peauger$^{3}$, 
J.~Plouin\\
}\textbf{CEA, Gif-sur-Yvette, France}
\end{center}
\vspace{-0.5cm}

\begin{center}{
P.~Bambade,
I.~Chaikovska,
R.~Chehab,
N.~Delerue, 
M.~Davier,
A.~Faus-Golfe, 
A.~Irles, 
W.~Kaabi,
F.~LeDiberder,
R.~P\"{o}schl,
D.~Zerwas\\
}\textbf{Laboratoire de l'Acc\'{e}l\'{e}rateur Lin\'{e}aire, Universit\'{e} de Paris-Sud XI, IN2P3/CNRS, Orsay, France}
\end{center}
\vspace{-0.5cm}

\begin{center}{
B.~Aimard,
G.~Balik,
J.-J.~Blaising, 
L.~Brunetti,
M.~Chefdeville, 
A.~Dominjon, 
C.~Drancourt,
N.~Geoffroy,
J.~Jacquemier,
A.~Jeremie,
Y.~Karyotakis,
J.M.~Nappa,
M.~Serluca, 
S.~Vilalte,
G.~Vouters\\
}\textbf{LAPP, Universit\'{e} de Savoie, IN2P3/CNRS, Annecy, France}
\end{center}
\vspace{-0.5cm}

\begin{center}{
A.~Bernhard, 
E.~Br\"{u}ndermann, 
S.~Casalbuoni, 
S.~Hillenbrand, 
J.~Gethmann, 
A.~Grau, 
E.~Huttel, 
A.-S.~M\"{u}ller, 
P.~Peiffer$^{4}$,  
I.~Peri\'{c},
D. Saez de Jauregui\\
}\textbf{Karlsruhe Institute of Technology (KIT), Karlsruhe, Germany} 
\end{center}
\vspace{-0.5cm}

\begin{center}{
L.~Emberger,
C.~Graf,
F.~Simon,
M.~Szalay,
N.~van~der~Kolk$^{5}$\\
}\textbf{Max-Planck-Institut f\"{u}r Physik, Munich, Germany}
\end{center}
\vspace{-0.5cm}

\begin{center}{
S.~Brass,
W.~Kilian\\
}\textbf{Department of Physics, University of Siegen, Siegen, Germany}
\end{center}
\vspace{-0.5cm}

\begin{center}{
T.~Alexopoulos,
T.~Apostolopoulos$^{6}$,  
E.N.~Gazis,
N.~Gazis,
V.~Kostopoulos$^{7}$,
S.~Kourkoulis\\
}\textbf{National Technical University of Athens, Athens, Greece}
\end{center}
\vspace{-0.5cm}

\begin{center}{
B.~Heilig\\
}\textbf{Department of Basic Geophysical Research, Mining and Geological Survey of Hungary, Tihany, Hungary}
\end{center}
\vspace{-0.5cm}

\begin{center}{
J.~Lichtenberger\\
}\textbf{Space Research Laboratory, E\"otv\"os Lor\'and University, Budapest, Hungary}
\end{center}
\vspace{-0.5cm}

\begin{center}{
P.~Shrivastava\\
}\textbf{Raja Ramanna Centre for Advanced Technology, Department of Atomic Energy, Indore, India}
\end{center}
\vspace{-0.5cm}

\begin{center}{
M.K.~Dayyani,
H.~Ghasem$^{1}$,
S.S.~Hajari,
H.~Shaker$^{1}$\\
}\textbf{The School of Particles and Accelerators, Institute for Research in Fundamental Sciences, Tehran, Iran}
\end{center}
\vspace{-0.5cm}

\begin{center}{
Y.~Ashkenazy,
I.~Popov, 
E.~Engelberg, 
A. Yashar\\ 
}\textbf{Racah Institute of Physics, Hebrew University of Jerusalem, Jerusalem, Israel}
\end{center}
\vspace{-0.5cm}

\begin{center}{
H.~Abramowicz,
Y.~Benhammou,
O.~Borysov,
M.~Borysova,
A.~Levy,
I.~Levy\\
}\textbf{Raymond \& Beverly Sackler School of Physics  \& Astronomy, Tel Aviv University, Tel Aviv, Israel}
\end{center}
\vspace{-0.5cm}

\begin{center}{
D.~Alesini,
M.~Bellaveglia,
B.~Buonomo,
A.~Cardelli,
M.~Diomede,
M.~Ferrario,
A.~Gallo,
A.~Ghigo,
A.~Giribono,
L.~Piersanti,
A.~Stella,
C.~Vaccarezza\\
}\textbf{INFN e Laboratori Nazionali di Frascati, Frascati, Italy}
\end{center}
\vspace{-0.5cm}

\begin{center}{
J.~de~Blas\\
}\textbf{Universit\`{a} di Padova and INFN, Padova, Italy}
\end{center}
\vspace{-0.5cm}

\begin{center}{
R.~Franceschini\\
}\textbf{Universit\`{a} degli Studi Roma Tre and INFN, Roma, Italy}
\end{center}
\vspace{-0.5cm}

\begin{center}{
G.~D'Auria, 
S.~Di Mitri\\
}\textbf{Elettra Sincrotrone Trieste, Trieste, Italy}
\end{center}
\vspace{-0.5cm}

\begin{center}{
T.~Abe, 
A.~Aryshev,
M.~Fukuda, 
K.~Furukawa, 
H.~Hayano, 
Y.~Higashi, 
T.~Higo,
K.~Kubo, 
S.~Kuroda, 
S.~Matsumoto, 
S.~Michizono, 
T.~Naito, 
T.~Okugi, 
T.~Shidara, 
T.~Tauchi, 
N.~Terunuma, 
J.~Urakawa, 
A.~Yamamoto$^{1}$\\
}\textbf{High Energy Accelerator Research Organization, KEK, Tsukuba, Japan}
\end{center}
\vspace{-0.5cm}

\begin{center}{
R.~Raboanary\\
}\textbf{University of Antananarivo, Antananarivo, Madagascar}
\end{center}
\vspace{-0.5cm}

\begin{center}{
O.J.~Luiten,
X.F.D.~Stragier\\
}\textbf{Eindhoven University of Technology, Eindhoven, Netherlands}
\end{center}
\vspace{-0.5cm}

\begin{center}{
R.~Hart,
H.~van der Graaf\\
}\textbf{Nikhef, Amsterdam, Netherlands}
\end{center}
\vspace{-0.5cm}

\begin{center}{
G.~Eigen\\
}\textbf{Department of Physics and Technology, University of Bergen, Bergen, Norway}
\end{center}
\vspace{-0.5cm}

\begin{center}{
E.~Adli$^{1}$,
C.A.~Lindstr{\o}m, 
R.~Lillest\o{}l,
L.~Malina$^1$,
J.~Pfingstner,
K.N.~Sjobak$^1$\\
}\textbf{University of Oslo, Oslo, Norway}
\end{center}
\vspace{-0.5cm}

\begin{center}{
A.~Ahmad, 
H.~Hoorani,
W.A.~Khan\\ 
}\textbf{National Centre for Physics, Islamabad, Pakistan}
\end{center}
\vspace{-0.5cm}

\begin{center}{
S.~Bugiel,
R.~Bugiel,
M.~Firlej,
T.A.~Fiutowski,
M.~Idzik,
J.~Moro\'{n},
K.P.~\'{S}wientek\\
}\textbf{AGH University of Science and Technology, Krakow, Poland}
\end{center}
\vspace{-0.5cm}

\begin{center}{
P.~Br\"{u}ckman~de~Renstrom, 
B.~Krupa,
M.~Kucharczyk,
T.~Lesiak,
B.~Pawlik,
P.~Sopicki,
B.~Turbiarz,
T.~Wojto\'{n},
L.K.~Zawiejski\\
}\textbf{Institute of Nuclear Physics, Polish Academy of Sciences, Krakow, Poland}
\end{center}
\vspace{-0.5cm}

\begin{center}{
J.~Kalinowski,
K.~Nowak,
A.F.~\.{Z}arnecki\\
}\textbf{Faculty of Physics, University of Warsaw, Warsaw, Poland}
\end{center}
\vspace{-0.5cm}

\begin{center}{
E.~Firu,
V.~Ghenescu,
A.T.~Neagu,
T.~Preda,
I. S. Zgura\\
}\textbf{Institute of Space Science, Bucharest, Romania}
\end{center}
\vspace{-0.5cm}

\begin{center}{
A.~Aloev,
N.~Azaryan,
I.~Boyko, 
J.~Budagov,
M.~Chizhov,
M.~Filippova,
V.~Glagolev,
A.~Gongadze,
S.~Grigoryan,
D.~Gudkov,
V.~Karjavine,
M.~Lyablin,
Yu.~Nefedov, 
A.~Olyunin$^{1}$,
A.~Rymbekova, 
A.~Samochkine,
A.~Sapronov, 
G.~Shelkov, 
G.~Shirkov,
V.~Soldatov,
E.~Solodko$^{1}$,
G.~Trubnikov,
I.~Tyapkin,
V.~Uzhinsky,
A.~Vorozhtov,
A.~Zhemchugov\\ 
}\textbf{Joint Institute for Nuclear Research, Dubna, Russia} 
\end{center}
\vspace{-0.5cm}

\begin{center}{
E.~Levichev,
N.~Mezentsev,
P.~Piminov,
D.~Shatilov,
P.~Vobly,
K.~Zolotarev\\
}\textbf{Budker Institute of Nuclear Physics, Novosibirsk, Russia}
\end{center}
\vspace{-0.5cm}

\begin{center}{
I.~Bo\v{z}ovi\'{c} Jelisav\v{c}i\'{c},
G.~Ka\v{c}arevi\'{c},
G.~Milutinovi\'{c} Dumbelovi\'{c},
M.~Pandurovi\'{c},
M.~Radulovi\'{c},
J. ~Stevanovi\'{c},
N.~Vukasinovi\'{c}\\
}\textbf{Vin\v{c}a Institute of Nuclear Sciences, University of Belgrade, Belgrade, Serbia}
\end{center}
\vspace{-0.5cm}

\begin{center}{
D.-H.~Lee\\
}\textbf{School of Space Research, Kyung Hee University, Yongin, Gyeonggi, South Korea}
\end{center}
\vspace{-0.5cm}

\begin{center}{
N.~Ayala, 
G.~Benedetti, 
T.~Guenzel, 
U.~Iriso,
Z.~Marti, 
F.~Perez,
M.~Pont\\
}\textbf{CELLS-ALBA, Barcelona, Spain}
\end{center}
\vspace{-0.5cm}

\begin{center}{
J.~Trenado\\
}\textbf{University of Barcelona, Barcelona, Spain}
\end{center}
\vspace{-0.5cm}

\begin{center}{
A.~Ruiz-Jimeno,
I.~Vila\\
}\textbf{IFCA, CSIC-Universidad de Cantabria, Santander, Spain}
\end{center}
\vspace{-0.5cm}

\begin{center}{
J.~Calero, 
M.~Dominguez, 
L.~Garcia-Tabares,
D.~Gavela,
D.~Lopez,
F.~Toral\\
}\textbf{Centro de Investigaciones Energ\'{e}ticas, Medioambientales y Tecnol\'{o}gicas (CIEMAT), Madrid, Spain}
\end{center}
\vspace{-0.5cm}

\begin{center}{

C.~Blanch Gutierrez,
M.~Boronat,
D.~Esperante$^{1}$,
E.~Fullana,
J.~Fuster,
 I.~Garc\'{\i}a, 
B.~Gimeno, 
P.~Gomis Lopez, 
D.~Gonz\'{a}lez, 
M.~Perell\'{o}, 
E.~Ros,
M.A.~Villarejo, 
A.~Vnuchenko, 
M.~Vos\\
}\textbf{Instituto de F\'{\i}sica Corpuscular (CSIC-UV), Valencia, Spain}
\end{center}
\vspace{-0.5cm}

\begin{center}{
R.~Brenner, 
Ch.~Borgmann,
T.~Ekel\"{o}f, 
M.~Jacewicz,  
M.~Olveg{\aa}rd, 
R.~Ruber, 
V.~Ziemann\\
}\textbf{Uppsala University, Uppsala, Sweden}
\end{center}
\vspace{-0.5cm}

\begin{center}{
D.~Aguglia,
J.~Alabau Gonzalvo,
M.~Alcaide Leon, 
N.~Alipour~Tehrani, 
M.~Anastasopoulos, 
A.~Andersson,
F.~Andrianala$^{8}$,
F.~Antoniou,
A.~Apyan, 
D.~Arominski$^{9}$,
K.~Artoos,
S.~Assly,
S.~Atieh,
C.~Baccigalupi, 
R.~Ballabriga~Sune,
D.~Banon~Caballero,
M.J.~Barnes,
J.~Barranco~Garcia,
A.~Bartalesi, 
J.~Bauche,
C.~Bayar, 
C.~Belver-Aguilar,
A.~Benot Morell$^{10}$,
M.~Bernardini, 
D.R.~Bett,
S.~Bettoni$^{11}$, 
M.~Bettencourt, 
B.~Bielawski, 
O.~Blanco Garcia,
N.~Blaskovic Kraljevic, 
B.~Bolzon$^{12}$,  
X.A.~Bonnin,
D.~Bozzini,
E.~Branger, 
E.~Brondolin,
O.~Brunner,
M.~Buckland$^{13}$,   
H.~Bursali, 
H.~Burkhardt,
D.~Caiazza, 
S.~Calatroni, 
M.~Campbell,
N.~Catalan~Lasheras,
B.~Cassany,
E.~Castro, 
R.H.~Cavaleiro Soares, 
M.~Cerqueira~Bastos,
A.~Cherif,
E.~Chevallay,
V.~Cilento$^{14}$,  
R.~Corsini, 
R.~Costa$^{15}$,  
B.~Cure, 
S.~Curt,
A.~Dal Gobbo, 
D.~Dannheim,
E.~Daskalaki, 
L.~Deacon, 
A.~Degiovanni, 
G.~De~Michele,
L.~De~Oliveira,
V.~Del~Pozo~Romano,
J.P.~Delahaye,
D.~Delikaris,
P.G.~Dias~de~Almeida$^{16}$,     
T.~Dobers,
S.~Doebert,
I.~Doytchinov, 
M.~Draper,
F.~Duarte~Ramos,
M.~Duquenne, 
N.~Egidos~Plaja,
K.~Elsener,
J.~Esberg,
M.~Esposito,
L.~Evans,
V.~Fedosseev,
P.~Ferracin,
A.~Fiergolski,
K.~Foraz,
A.~Fowler,
F.~Friebel,
J-F.~Fuchs,
A.~Gaddi,
D.~Gamba, 
L.~Garcia~Fajardo$^{17}$,    
H.~Garcia~Morales,
C.~Garion,
M.~Gasior, 
L.~Gatignon,
J-C.~Gayde,
A.~Gerbershagen, 
H.~Gerwig,
G.~Giambelli, 
A.~Gilardi, 
A.N.~Goldblatt,
S.~Gonzalez~Anton, 
C.~Grefe$^{18}$,    
A.~Grudiev,
H.~Guerin, 
F.G.~Guillot-Vignot,
M.L.~Gutt-Mostowy,
M.~Hein Lutz, 
C.~Hessler,
J.K.~Holma,
E.B.~Holzer,
M.~Hourican,
D.~Hynds$^{19}$, 
E.~Ikarios,  
Y.~Inntjore~Levinsen,
S.~Janssens, 
A.~Jeff, 
E.~Jensen,
M.~Jonker,
S.W.~Kamugasa, 
M.~Kastriotou,
J.M.K.~Kemppinen,
V.~Khan, 
R.B.~Kieffer,
W.~Klempt,
N.~Kokkinis,
I.~Kossyvakis, 
Z.~Kostka, 
A.~Korsback,
E.~Koukovini~Platia,
J.W.~Kovermann,
C-I.~Kozsar,
I.~Kremastiotis$^{20}$,
J.~Kr\"{o}ger$^{21}$,
S.~Kulis,
A.~Latina,
F.~Leaux,
P.~Lebrun, %
T.~Lefevre,
E.~Leogrande,
L.~Linssen,
X.~Liu, 
X.~Llopart~Cudie,
S.~Magnoni,  
C.~Maidana, 
A.A.~Maier,
H.~Mainaud~Durand,
S.~Mallows, 
E.~Manosperti,
C.~Marelli, 
E.~Marin~Lacoma,
S.~Marsh, 
R.~Martin,
I.~Martini, 
M.~Martyanov, 
S.~Mazzoni,
G.~Mcmonagle,
L.M.~Mether,
C.~Meynier, 
M.~Modena,
A.~Moilanen, 
R.~Mondello, 
P.B.~Moniz Cabral,
N.~Mouriz Irazabal, 
M.~Munker,
T.~Muranaka,
J.~Nadenau, 
J.G.~Navarro, 
J.L.~Navarro Quirante, 
E.~Nebo~ Del~Busto,
N.~Nikiforou$^{22}$,
P.~Ninin,
M.~Nonis,
D.~Nisbet,
F.X.~Nuiry,
A.~N\"{u}rnberg$^{23}$,
J.~\"{O}gren, 
J.~Osborne,
A.C.~Ouniche, 
R.~Pan$^{24}$, 
S.~Papadopoulou,
Y.~Papaphilippou,
G.~Paraskaki, 
A.~Pastushenko$^{10}$, 
A.~Passarelli,
M.~Patecki,
L.~Pazdera,
D.~Pellegrini,
K.~Pepitone,
E.~Perez~Codina,
A.~Perez~Fontenla,
T.H.B.~Persson,
M.~Petri\v{c}$^{25}$,
S.~Pitman, 
F.~Pitters$^{26}$,
S.~Pittet,
F.~Plassard,
D.~Popescu, 
T.~Quast$^{27}$,
R.~Rajamak,
S.~Redford$^{11}$,
L.~Remandet, 
Y.~Renier$^{24}$,
S.F.~Rey,
O.~Rey~Orozco, 
G.~Riddone,
E.~Rodriguez~Castro,
P.~Roloff,    %
C.~Rossi,
F.~Rossi, 
V.~Rude,
I.~Ruehl, 
G.~Rumolo,
A.~Sailer,
J.~Sandomierski, 
E.~Santin,
C.~Sanz, 
J.~Sauza Bedolla, 
U.~Schnoor,
H.~Schmickler,
D.~Schulte,
E.~Senes, 
C.~Serpico,
G.~Severino, 
N.~Shipman,
E.~Sicking,
R.~Simoniello$^{28}$,
P.K.~Skowronski,
P.~Sobrino~Mompean,
L.~Soby,
P.~Sollander,
A.~Solodko, 
M.P.~Sosin,
S.~Spannagel,
S.~Sroka,
S.~Stapnes,
G.~Sterbini,
G.~Stern, 
R.~Str\"{o}m,
M.J.~Stuart,
I.~Syratchev,
K.~Szypula, 
F.~Tecker,
P.A.~Thonet,
P.~Thrane, 
L.~Timeo,
M.~Tiirakari, 
R.~Tomas~Garcia,
C.I.~Tomoiaga, 
P.~Valerio$^{29}$,
T.~Va\v{n}\'{a}t,
A.L.~Vamvakas,
J.~Van~Hoorne,  
O.~Viazlo,
M.~Vicente~Barreto~Pinto$^{30}$,
N.~Vitoratou, 
V.~Vlachakis, 
M.A.~Weber,
R.~Wegner,
M.~Wendt,
M.~Widorski,
O.E.~Williams, 
M.~Williams$^{31}$,
B.~Woolley,
W.~Wuensch,
A.~Wulzer,
J.~Uythoven,
A.~Xydou, 
R.~Yang,
A.~Zelios, 
Y.~Zhao$^{32}$, 
P.~Zisopoulos\\
}\textbf{CERN, Geneva, Switzerland}
\end{center}
\vspace{-0.5cm}

\begin{center}{
M.~Benoit,
D~M~S~Sultan\\
}\textbf{D\'{e}partement de Physique Nucl\'{e}aire et Corpusculaire (DPNC), Universit\'{e} de Gen\`{e}ve, Gen\`{e}ve, Switzerland}
\end{center}
\vspace{-0.5cm}

\begin{center}{
F.~Riva$^{1}$\\
}\textbf{D\'{e}partement de Physique Th\'{e}orique, Universit\'{e} de Gen\`{e}ve, Gen\`{e}ve, Switzerland}
\end{center}
\vspace{-0.5cm}

\begin{center}{
M.~Bopp,
H.H.~Braun,
P.~Craievich, 
M.~Dehler,
T.~Garvey,
M.~Pedrozzi, 
J.Y.~Raguin,
L.~Rivkin$^{33}$,
R.~Zennaro\\
}\textbf{Paul Scherrer Institut, Villigen, Switzerland}
\end{center}
\vspace{-0.5cm}

\begin{center}{
S.~Guillaume,  
M.~Rothacher\\  
}\textbf{ETH Zurich, Institute of Geodesy and Photogrammetry, Zurich, Switzerland}
\end{center}
\vspace{-0.5cm}

\begin{center}{
A.~Aksoy,
Z.~Nergiz$^{34}$,
\"{O}.~Yavas\\
}\textbf{Ankara University, Ankara, Turkey}
\end{center}
\vspace{-0.5cm}

\begin{center}{
H.~Denizli,
U.~Keskin, 
K.~Y.~Oyulmaz,
A.~Senol\\
}\textbf{Department of Physics, Abant \.{I}zzet Baysal University, Bolu, Turkey}
\end{center}
\vspace{-0.5cm}

\begin{center}{
A.K.~Ciftci\\
}\textbf{Izmir University of Economics, Izmir, Turkey}
\end{center}
\vspace{-0.5cm}

\begin{center}{
V.~Baturin,
O.~Karpenko, 
R.~Kholodov,
O.~Lebed, 
S.~Lebedynskyi,
S.~Mordyk,
I.~Musienko, 
Ia.~Profatilova, 
V.~Storizhko\\
}\textbf{Institute of Applied Physics, National Academy of Sciences of Ukraine, Sumy, Ukraine}
\end{center}
\vspace{-0.5cm}

\begin{center}{
R.R.~Bosley,
T.~Price,
M.F.~Watson,
N.K.~Watson,
A.G.~Winter\\
}\textbf{University of Birmingham, Birmingham, United Kingdom}
\end{center}
 
\begin{center}{
J.~Goldstein\\
}\textbf{University of Bristol, Bristol, United Kingdom}
\end{center}
\vspace{-0.5cm}

\begin{center}{
S.~Green,
J.S.~Marshall$^{35}$,
M.A.~Thomson,
B.~Xu,
T.~You$^{36}$\\
}\textbf{Cavendish Laboratory, University of Cambridge, Cambridge, United Kingdom}
\end{center}
\vspace{-0.5cm}

\begin{center}{
W.A.~Gillespie\\
}\textbf{University of Dundee, Dundee, United Kingdom}
\end{center}
\vspace{-0.5cm}

\begin{center}{
M.~Spannowsky\\
}\textbf{Department of Physics, Durham University, Durham, United Kingdom}
\end{center}
\vspace{-0.5cm}

\begin{center}{
C.~Beggan\\
}\textbf{British Geological Survey, Edinburgh, United Kingdom}
\end{center}
\vspace{-0.5cm}

\begin{center}{
V.~Martin,
Y.~Zhang\\
}\textbf{University of Edinburgh, Edinburgh, United Kingdom}
\end{center}
\vspace{-0.5cm}

\begin{center}{
D.~Protopopescu,
A.~Robson$^{1}$\\
}\textbf{University of Glasgow, Glasgow, United Kingdom}
\end{center}
\vspace{-0.5cm}

\begin{center}{
R.J.~Apsimon$^{37}$,
I.~Bailey$^{38}$,
G.C.~Burt$^{37}$,
A.C.~Dexter$^{37}$,
A.V.~Edwards$^{37}$, 
V.~Hill$^{37}$, 
S.~Jamison, 
W.L.~Millar$^{37}$, 
K.~Papke$^{37}$\\  
}\textbf{Lancaster University, Lancaster, United Kingdom}
\end{center}
\vspace{-0.5cm}

\begin{center}{
G.~Casse, 
J.~Vossebeld\\
}\textbf{University of Liverpool, Liverpool, United Kingdom}
\end{center}
\vspace{-0.5cm}

\begin{center}{
T.~Aumeyr, 
M.~Bergamaschi$^{1}$, 
L.~Bobb$^{38}$, 
A.~Bosco, 
S.~Boogert, 
G.~Boorman, 
F.~Cullinan, 
S.~Gibson, 
P.~Karataev,
K.~Kruchinin, 
K.~Lekomtsev, 
A.~Lyapin, 
L.~Nevay, 
W.~Shields, 
J.~Snuverink, 
J.~Towler, 
E.~Yamakawa\\ 
}\textbf{The John Adams Institute for Accelerator Science, Royal Holloway, University of London, Egham, United Kingdom}
\end{center}
\vspace{-0.5cm}

\begin{center}{
V.~Boisvert,
S.~West\\
}\textbf{Royal Holloway, University of London, Egham, United Kingdom}
\end{center}
\vspace{-0.5cm}

\begin{center}{
R.~Jones,
N.~Joshi\\
}\textbf{University of Manchester, Manchester, United Kingdom}
\end{center}
\vspace{-0.5cm}

\begin{center}{

D.~Bett, 
R.M.~Bodenstein$^{1}$, 
T.~Bromwich,
P.N.~Burrows$^{1}$, 
G.B.~Christian$^{38}$, 
C.~Gohil$^{1}$, 
P.~Korysko$^{1}$, 
J.~Paszkiewicz$^{1}$, 
C.~Perry,
R.~Ramjiawan, 
J.~Roberts\\ 
}\textbf{John Adams Institute, Department of Physics, University of Oxford, Oxford, United Kingdom}
\end{center}
\vspace{-0.5cm}

\begin{center}{
T.~Coates,
F.~Salvatore\\
}\textbf{University of Sussex, Brighton, United Kingdom}
\end{center}
\vspace{-0.5cm}

\begin{center}{
A.~Bainbridge$^{37}$, 
J.A.~Clarke$^{37}$,
N.~Krumpa,   
B.J.A.~Shepherd$^{37}$,
D.~Walsh$^{3}$\\
}\textbf{STFC Daresbury Laboratory, Warrington, United Kingdom}
\end{center}
\vspace{-0.5cm}

\begin{center}{
S.~Chekanov, 
M.~Demarteau, 
W.~Gai, 
W.~Liu, 
J.~Metcalfe, 
J.~Power, 
J.~Repond, 
H.~Weerts, 
L.~Xia,     
J.~Zhang\\ 
}\textbf{Argonne National Laboratory, Argonne, USA}
\end{center}
\vspace{-0.5cm}

\begin{center}{
J.~Zupan\\
}\textbf{Department of Physics, University of Cincinnati, Cincinnati, OH, USA}
\end{center}
\vspace{-0.5cm}

\begin{center}{
J.D.~Wells,
Z.~Zhang\\
}\textbf{Physics Department, University of Michigan, Ann Arbor, MI, USA}
\end{center}
\vspace{-0.5cm}

\begin{center}{
C.~Adolphsen, 
T.~Barklow, 
V.~Dolgashev, 
M.~Franzi, 
N.~Graf, 
J.~Hewett, 
M.~Kemp, 
O.~Kononenko, 
T.~Markiewicz, 
K.~Moffeit, 
J.~Neilson, 
Y.~Nosochkov,
M.~Oriunno, 
N.~Phinney, 
T.~Rizzo, 
S.~Tantawi, 
J.~Wang, 
B.~Weatherford, 
G.~White, 
M.~Woodley\\
}\textbf{SLAC National Accelerator Laboratory, Menlo Park, USA}
\end{center}
\vspace{-0.5cm}

\begin{flushleft}{
{$^{1}$}Also at CERN, Geneva, Switzerland\\
{$^{2}$}Also at Department of Physics, University of Helsinki, Helsinki, Finland\\
{$^{3}$}Now at CERN, Geneva, Switzerland\\
{$^{4}$}Now at Johannes-Gutenberg University, Mainz, Germany\\
{$^{5}$}Now at Nikhef / Utrecht University, Amsterdam / Utrecht, The Netherlands\\
{$^{6}$}Also at Department of Informatics, Athens University of Business and Economics, Athens, Greece \\
{$^{7}$}Also at University of Patras, Patras, Greece\\
{$^{8}$}Also at University of Antananarivo, Antananarivo, Madagascar\\
{$^{9}$}Also at Warsaw University of Technology, Warsaw, Poland\\
{$^{10}$}Also at IFIC, Valencia, Spain\\
{$^{11}$}Now at Paul Scherrer Institute, Villigen, Switzerland\\
{$^{12}$}Now at CEA, Gif-sur-Yvette, France\\   
{$^{13}$}Also at University of Liverpool, United Kingdom\\
{$^{14}$}Also at LAL, Orsay, France\\
{$^{15}$}Also at Uppsala University, Uppsala, Sweden\\ 
{$^{16}$}Also at IFCA, CSIC-Universidad de Cantabria, Santander, Spain\\
{$^{17}$}Now at LBNL, Berkeley CA, USA\\
{$^{18}$}Now at University of Bonn, Bonn, Germany\\
{$^{19}$}Now at Nikhef, Amsterdam, The Netherlands\\
{$^{20}$}Also at KIT, Karlsruhe, Germany\\
{$^{21}$}Also at Ruprecht-Karls-Universit\"{a}t Heidelberg, Germany\\
{$^{22}$}Now at University of Texas, Austin, USA\\
{$^{23}$}Now at Karlsruhe Institute of Technology, Karlsruhe, Germany\\
{$^{24}$}Now at DESY, Zeuthen, Germany\\
{$^{25}$}Also at J.\ Stefan Institute, Ljubljana, Slovenia\\
{$^{26}$}Also at Vienna University of Technology, Vienna, Austria\\
{$^{27}$}Also at RWTH Aachen University, Aachen, Germany\\
{$^{28}$}Now at Johannes Gutenberg Universit\"{a}t, Mainz, Germany\\
{$^{29}$}Now at D\'{e}partement de Physique Nucl\'{e}aire et Corpusculaire (DPNC), Universit\'{e} de Gen\`{e}ve, Geneva, Switzerland\\
{$^{30}$}Also at D\'{e}partement de Physique Nucl\'{e}aire et Corpusculaire (DPNC), Universit\'{e} de Gen\`{e}ve, Geneva, Switzerland\\
{$^{31}$}Also at University of Glasgow, Glasgow, United Kingdom\\
{$^{32}$}Also at Shandong University, Jinan, China\\
{$^{33}$}Also at EPFL, Lausanne, Switzerland\\
{$^{34}$}Also at Omer Halis Demir University, Nigde, Turkey\\
{$^{35}$}Now at University of Warwick, Coventry, United Kingdom\\
{$^{36}$}Also at DAMTP, University of Cambridge, Cambridge, United Kingdom\\
{$^{37}$}Also at The Cockcroft Institute, Daresbury, United Kingdom\\
{$^{38}$}Now at Diamond Light Source, Harwell, United Kingdom}
\end{flushleft}
\end{multicols}
}

\end{document}